\documentclass[twoside,twocolumn,9pt]{article}

\usepackage{extsizes}
\usepackage[super,sort&compress,comma]{natbib} 
\usepackage[left=1.5cm, right=1.5cm, top=1.785cm, bottom=2.0cm]{geometry}
\usepackage{graphicx} 
\usepackage{float}
\usepackage[T1]{fontenc}
\usepackage[utf8]{inputenc}
\usepackage[compact]{titlesec}
\usepackage{amssymb}
\usepackage{amsmath}

\begin{document}

\noindent\LARGE{\textbf{Theory of electronic resonances: \newline
Fundamental aspects and recent advances}} \\

 \noindent\large{Thomas-C. Jagau $^{a}$} \\[0.5cm]

 \noindent\normalsize{
Electronic resonances are states that are unstable towards loss of electrons. They play critical 
roles in high-energy environments across chemistry, physics, and biology but are also relevant 
to processes under ambient conditions that involve unbound electrons. This feature article 
focuses on complex-variable techniques such as complex scaling and complex absorbing 
potentials that afford a treatment of electronic resonances in terms of discrete square-integrable 
eigenstates of non-Hermitian Hamiltonians with complex energy. Fundamental aspects of these 
techniques as well their integration into molecular electronic-structure theory are discussed and 
an overview of some recent developments is given: analytic gradient theory for electronic 
resonances, the application of rank-reduction techniques and quantum embedding to them, 
as well as approaches for evaluating partial decay widths.} \\
\vspace{0.5cm} 

\renewcommand*\rmdefault{bch}\normalfont\upshape
\rmfamily
\section*{}
\vspace{-1cm}


\footnotetext{\textit{$^{a}$~Department of Chemistry, KU Leuven, Celestijnenlaan 200F, 
B-3001 Leuven, Belgium. 
Tel: +32 16 32 7939; E-mail: thomas.jagau@kuleuven.be}}





\section{Introduction} \label{sec:intro}
It is usually assumed in quantum chemistry without further discussion that the Hamiltonian is 
Hermitian and the energies, which are obtained as eigenvalues of the Hamiltonian, thus real. 
However, already in 1928, George Gamow recognized that $\alpha$ decay of radioactive nuclei 
can be modeled in terms of a non-Hermitian Hamiltonian and complex-valued energies, whose 
imaginary parts are interpreted as decay widths.\cite{gamov28}

Whereas $\alpha$ decay has always been of central importance for nuclear physics, and complex 
energies hence occur quite frequently in this field,\cite{myo20} corresponding electronic decay 
processes played a subordinate role for chemistry for many years. Complex energies were 
considered to be an exotic phenomenon and were discussed mostly as an 
unwanted byproduct of some electronic-structure models, notably truncated coupled-cluster 
methods.\cite{haettig05,koehn07,kjonstad17a,thomas21} However, driven by the growing 
relevance of processes involving unbound electrons, the importance of electronic decay for 
chemistry has increased in recent years and is expected to continue increasing. State-of-the-art 
experimental techniques make it possible to create, in a controlled manner, environments where 
selected electrons are no longer bound to the nuclei. For example, core vacancies produced by 
X rays,\cite{xraybook,norman18,zimmermann20} temporary anions obtained by attachment of 
slow electrons,\cite{simons11,herbert15,alizadeh15} and molecules exposed to quasistatic laser 
fields\cite{scrinzi05,gallmann12} all undergo electronic decay. 

The subject of this feature article is the quantum-chemical treatment of states that govern 
electronic decay processes, so-called electronic resonances, in terms of 
complex energies.\cite{nhqmbook,reinhardt82,jagau17} The resonance 
phenomenon as such is, however, relevant to other branches of science as well: $\alpha$ 
decay of radioactive nuclei, predissociation of van der Waals complexes, as well as leaky 
modes of optical waveguides all represent examples. An excellent overview is available 
from Ref. \citenum{nhqmbook}. A recent perspective on rotational-vibrational resonance 
states and their role in chemistry is available from Ref. \citenum{csaszar20}. Furthermore, 
non-Hermitian Hamiltonians have been used to model molecular electronics.\cite{baer03,
ernzerhof06}

Although substantial methodological progress has been made in recent years, \textit{ab initio} 
modeling involving electronic decay remains very challenging. Except for special cases in 
which a resonance can be easily decoupled from the embedding continuum, for example, 
a core-vacant state by means of the core-valence separation,\cite{cederbaum80} coupling to 
the continuum needs to be considered but quantum-chemical methods designed for bound 
states, i.e., the discrete part of the Hamiltonian's spectrum, cannot accomplish this without time 
propagation.\cite{nhqmbook,jagau17} While a complete description of resonances and decay 
phenomena is obtained by solving the time-dependent Schr\"odinger equation, these approaches 
entail tremendous computational cost and remain limited to very small systems. 

This makes a time-independent treatment of resonances desirable. Several approaches have 
been developed for this purpose: A pragmatic approach consists in extrapolating results from 
bound-state calculations using various stabilization\cite{hazi70,nestmann85,mandelshtam94} 
or analytic continuation\cite{simons81b,mccurdy83,frey86,horacek10,horacek15,sommerfeld15b,
white17b} methods. A more rigorous description is offered by scattering theory where one 
imposes scattering boundary conditions on the solution of the Schr\"odinger equation.\cite{
nhqmbook,taylor72,domcke91} This works very well for atoms and model systems but the 
integration into molecular electronic-structure theory is challenging. Important 
approaches for electron-molecule collisions based on scattering theory include the R-matrix 
method,\cite{tennyson10} the Schwinger multichannel method,\cite{dacosta15} and the discrete 
momentum representation method.\cite{lane80} Overviews of recent developments in this field 
are available, for example, from Refs. \citenum{ingolfsson19,gorfinkiel20,masin20}. In the 
following, the treatment of molecular electronic resonances in terms of scattering theory is 
not discussed further.

To certain types of resonances, the theory by Fano and Feshbach\cite{fano61,feshbach62,
domcke91,averbukh05,kolorenc20} can be applied, which treats a resonance as a bound 
state superimposed on the continuum. Here, a projection-operator formalism is used to 
divide the Hilbert space into a bound and a continuum part. A distinct advantage is that 
standard quantum-chemical methods can be used but the critical step consists in the 
definition of the projector. Also, additional steps need to be taken to extract information 
on the decay process from the discretized representation of the electronic continuum 
obtained in such calculations. This can be done implicitly, for example, by means of 
Stieltjes imaging,\cite{langhoff74,carravetta87} or alternatively by modeling the wave 
function of the outgoing electron explicitly.\cite{zaehringer92,inhester12,skomorowski21}

A further, more general approach relies on recasting electronic resonances as $\mathcal{L}^2$ 
integrable wave functions by means of complex-variable techniques,\cite{nhqmbook,jagau17} 
in particular complex scaling,\cite{aguilar71,balslev71,simon72,moiseyev98} complex-scaled 
basis functions,\cite{mccurdy78,moiseyev79} and complex absorbing potentials.\cite{jolicard85,
riss93} No explicit treatment of the electronic continuum is needed here, which offers the 
possibility to apply methods and concepts from bound-state quantum chemistry. This 
affords a description of decaying states in terms that are familiar to quantum chemists 
such as molecular orbitals and potential energy surfaces.

The latter methods, which revolve around non-Hermitian Hamiltonians with complex energy 
eigenvalues, are the focus of the present feature article. The remainder of the article is structured 
as follows: In Sec. \ref{sec:res}, an overview of different types of electronic resonances and their 
relevance for chemistry is given. Sec. \ref{sec:cv} summarizes the theoretical foundations of 
complex-variable techniques and discusses practical aspects of their implementation into 
quantum-chemical program packages. Sec. \ref{sec:dev} showcases a number of recent 
methodological developments together with exemplary applications, while Sec. \ref{sec:conc} 
presents some general conclusions and speculations about future developments in the field. 


\section{Electronic resonances} \label{sec:res}
Many types of electronic decay processes exist and the electronic structure of the corresponding 
resonance states is as diverse as those of bound states. In the following, an overview of several 
types of states is given: Autodetaching anions, core-vacant states, and Stark resonances formed 
in static electric fields. In addition, the distinction between shape and Feshbach resonances is 
discussed.

\subsection{Autodetaching anions} \label{sec:anion}
If one considers a neutral molecule, the corresponding anion is stable only if it is lower in energy, 
that is, if the electron affinity is positive. If this is not the case, it does not imply that a free electron 
is necessarily scattered off elastically. Instead, a temporary anion with a lifetime in the range of 
femtoseconds to milliseconds can form in an electron-molecule collision.\cite{simons08,simons11,
herbert15,jagau17,nhqmbook} These species are ubiquitous; in fact, most closed-shell molecules 
do not form stable but only temporary anions in gas phase.\cite{simons08,herbert15}

Temporary anions play a central role for dissociative electron attachment (DEA) and related 
electron-induced reactions.\cite{herbert15,fabrikant17,ingolfsson19} Often, reaction barriers 
that are insurmountable on the potential energy surface of the neutral species can be 
overcome in the presence of catalytic electrons with a few electron volts of kinetic energy. 
This is widely exploited in organic synthesis using electronically bound anions\cite{studer14} 
but can also involve temporary anions. The latter are, for example, relevant to plasmonic 
catalysis,\cite{aslam18} nitrogen fixation using cold plasma,\cite{li18} and radiation-induced 
damage to living tissue.\cite{boudaiffa00,alizadeh15} 

A further type of resonance are excited states of stable closed-shell anions.\cite{simons08,
simons11,herbert15} Species such as F$^-$, OH$^-$, and CN$^-$ only rarely 
support bound excited states; electronic excitation usually entails electron detachment. The 
corresponding photodetachment spectra often feature distinct fingerprints of metastable 
states.\cite{jagau15,lyle18,lyle19} Also, dianions including common species such as O$^{2-}$, 
SO$_4^{2-}$, and CO$_3^{2-}$ are almost never stable against electron loss in gas phase 
and exist as bound electronic states only if a solvation shell or some other environment is 
present.\cite{dreuw02} 

\begin{figure} \centering
\includegraphics[scale=0.5]{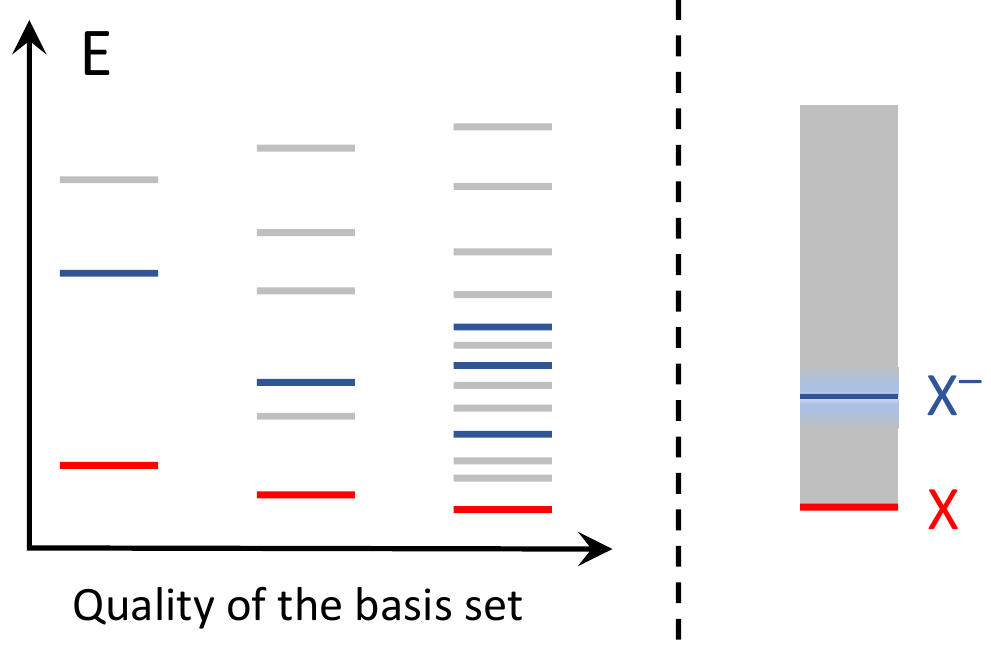}
\caption{Description of an autodetaching anion X$^-$ in a standard electronic-structure 
calculation. Left: When the size of the basis set is increased, the description of the continuum 
(gray) improves and the resonance (blue) cannot be associated with a single discrete state 
anymore. Right: In a complete basis set, the resonance can be associated with a peak in 
the density of continuum states. The onset of the electronic continuum is shown in red. 
Reproduced with permission from Ref. \citenum{jagau20}.}
\label{fig:basis}
\end{figure}

All these anions have in common that they are beyond the reach of standard quantum-chemical 
methods.\cite{jagau17} This is illustrated in Fig. \ref{fig:basis}. Since one always uses a finite 
number of basis functions in an actual computation, the continuum is discretized. If the basis 
is small, usable approximations of resonances as bound states may be obtained in some cases 
because no basis function can describe the coupling to the continuum and, consequently, the 
decay process. However, this is a crude and uncontrolled approximation: If the size of the basis is 
increased, the representation of the continuum improves and the resonance cannot be associated 
with a single state anymore. It \textit{dissolves} in the continuum.\cite{jagau17,bravaya13,jagau16b} 
An electronic-structure calculation will collapse to a continuum-like solution where one electron 
is as far away from the molecule as the basis set permits. In the limit of an infinite basis set 
(see right-hand side of Fig. \ref{fig:basis}), where the continuum is represented properly, one 
observes an increased density of states in a certain energy range. Since the position and width 
of such a peak can be associated with the energy and inverse lifetime of a resonance, the terms 
\textit{resonance position} and \textit{resonance width} have been coined.\cite{nhqmbook,
reinhardt82,jagau17} 

\subsection{Core-vacant states} \label{sec:core}
Electronic resonances are also encountered when moving beyond anions with the difference 
that the decay of neutral or cationic species is termed autoionization instead of autodetachment. 
Neutral states above the first ionization threshold are also called superexcited states. An 
important subgroup of autoionizing resonances consists in core-vacant states, which are 
created by interaction with X-rays in various spectroscopies.\cite{xraybook,norman18,
zimmermann20} These techniques typically involve photon energies larger than 200 eV 
up to 1000 eV so that the resulting states are located at much higher energies than the 
temporary anions from Sec. \ref{sec:anion}. 

\begin{figure} \centering
\includegraphics[scale=0.34]{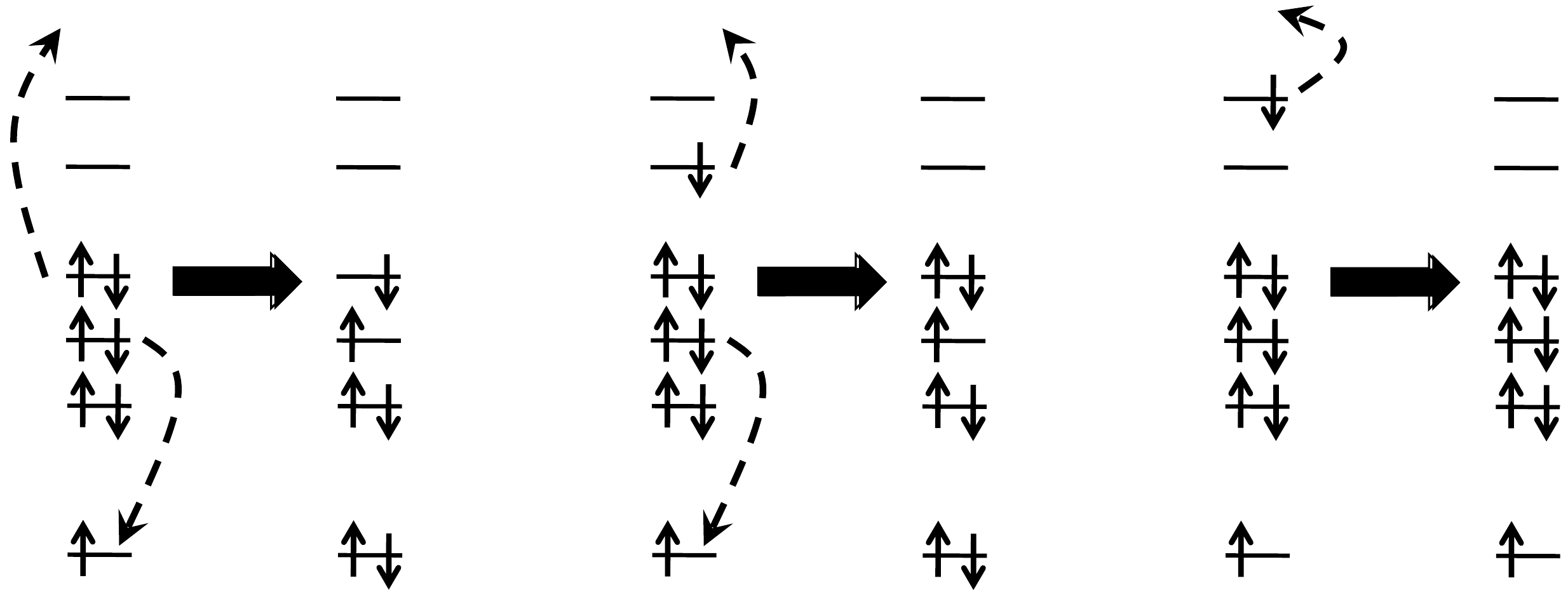}
\caption{Non-radiative decay of core-vacant states: In Auger 
decay depicted in the left panel, a core-ionized cationic state decays into a 
dicationic state. In resonant Auger decay depicted in the middle, a core-excited neutral state 
decays into a cationic state. There are also core-excited states above the 
respective core-ionization threshold that can undergo a one-electron decay process in which 
the core hole is not filled as shown in the right panel.}
\label{fig:auger}
\end{figure}

Core-vacant states are subject to the Auger-Meitner effect,\cite{meitner22,auger23} in which 
the core vacancy is filled while a second electron is emitted into the ionization continuum. 
Different variants of Auger decay and the corresponding decay channels can be identified 
in Auger electron spectroscopy\cite{xraybook} by measuring the kinetic energy of the emitted 
electrons. Several important non-radiative decay processes that can follow
initial core ionization or core excitation,\cite{brown80,piancastelli87,kempgens99,armen00} 
are depicted in Fig. \ref{fig:auger}. Further related processes besides those in Fig. \ref{fig:auger} 
include double Auger decay,\cite{carlson65} where two electrons are simultaneously emitted, 
and various shake-up and shake-off mechanisms,\cite{koerber66,hotokka84,colle90} 
where an additional valence electron is ionized or excited, respectively. 
It is also common that the target states of Auger decay undergo further decay resulting 
in so-called Auger cascades.\cite{xraybook} 

An important common aspect of core-vacant states is that they can be modeled as bound 
states with much better accuracy than other types of resonances. In particular, it is possible 
to project out the ionization continuum by means of the core-valence separation.\cite{
cederbaum80,coriani15,zheng19,vidal19} This is done by removing those configurations 
from the Hilbert space that describe the Auger decay. Notably, such decoupling is not 
possible for core-excited states above the respective core-ionization threshold (right panel 
of Fig. \ref{fig:auger}).

Related to Auger decay are non-local processes in which the emitted electron does not 
stem from the atom or molecule in which the core hole was located. The prime example 
is intermolecular Coulombic decay (ICD),\cite{cederbaum97} but there are further flavors 
such as electron-transfer mediated decay\cite{zobeley01} and interatomic Coulombic 
electron capture.\cite{gokhberg09} ICD and related processes are possible at considerably 
lower energies than Auger decay and thus presumed to be more widespread.\cite{jahnke20} 
Also, the efficiency of ICD increases in the presence of many neighboring molecules that 
can be ionized, which further contributes to its relevance in complex systems. It has even 
been claimed that \textit{ICD is everywhere}.\cite{ouchi11}

\subsection{Quasistatic ionization} \label{sec:quasi}
A further type of electronic resonance, which is not connected to autodetachment or autoionization, 
arises when atoms or molecules are exposed to intense laser fields that are comparable in strength 
to the intramolecular forces. Under such conditions, quasistatic ionization takes place and bound 
states are turned into Stark resonances.\cite{nhqmbook,reinhardt82} This process underlies many 
phenomena observed in strong laser fields, in particular high-harmonic generation.\cite{scrinzi05,
gallmann12} While a comprehensive discussion of atoms and molecules in laser fields is beyond 
the scope of this article, a brief account of quasistatic ionization is given in the following. 

\begin{figure}
\includegraphics[scale=0.5]{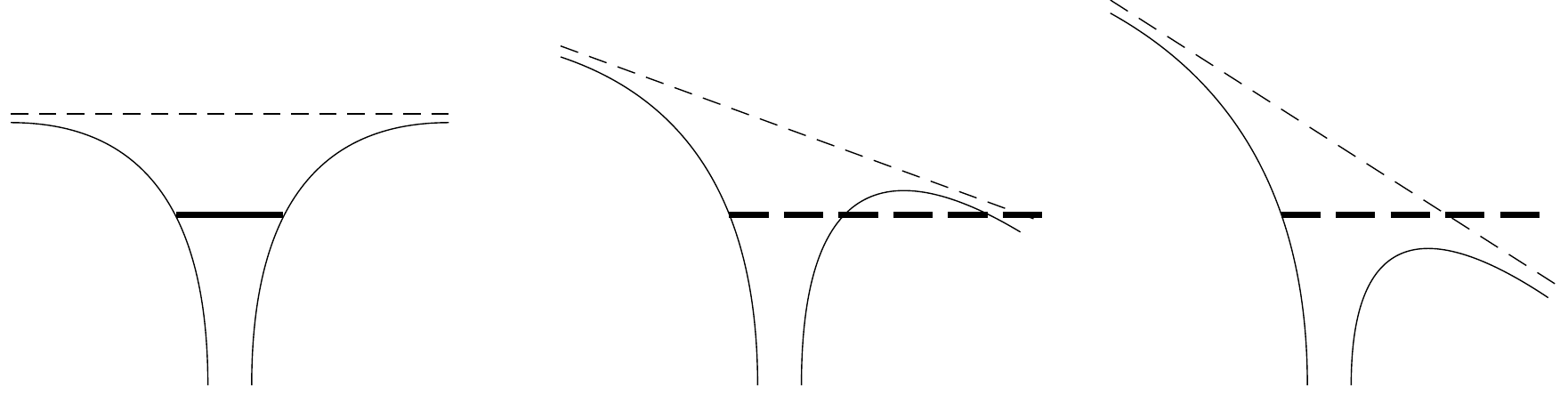}
\caption{Distortion of a Coulombic potential by a static electric field in one dimension. A stable 
energy level (left) becomes metastable in the presence of a field (middle). At higher field 
strengths, barrier-suppression ionization is possible (right). Reproduced with permission from 
Ref. \citenum{jagau16}.}
\label{fig:stark}
\end{figure}

One can distinguish alternating current (ac) Stark resonances formed in time-dependent electric 
fields and direct current (dc) Stark resonances formed in static electric fields. The formation of dc 
Stark resonances can be understood in terms of Fig. \ref{fig:stark}: Owing to the distortion of the 
potential, electrons can leave the system by tunneling. At even higher field strengths, electrons can 
leave above the barrier, which is termed barrier-suppression ionization.\cite{scrinzi05,scrinzi99} 
Strictly speaking, tunnel ionization is already possible at infinitesimally low field strengths meaning 
that the Hamiltonian of a molecule in the presence of an external electric field never supports any 
bound states.\cite{herbst79,herbst81,caliceti07} 

This effect can be ignored if the field strength is low enough and a treatment of light-matter 
interaction in terms of response properties is possible.\cite{helgaker12} However, the radius 
of convergence of a perturbative expansion in powers of the field strength is determined by 
the tunnel ionization rate.\cite{caliceti07} At higher field strengths where decay widths are 
significant; a treatment of the external field as perturbation is not valid. 

To determine under what conditions quasistatic ionization takes actually place, not only the 
field strength but also the frequency of a laser need to be considered.\cite{scrinzi05,gallmann12,
reiss08} For small molecules in the electronic ground state, field strengths of the order of $10^{-4}$ 
to 1 a.u. are usually of interest. This exceeds field strengths typical for photochemistry by orders 
of magnitudes, the realization of such conditions is, however, no problem in modern laser 
experiments.\cite{scrinzi05,gallmann12} By means of Keldysh's adiabaticity parameter, different 
ionization mechanisms can be distinguished.\cite{keldysh65} This relates the tunneling time to 
a wave period and provides thus an estimate if the ionization can be thought of as static process 
as depicted in Fig. \ref{fig:stark}. If this is the case, the time dependence of the field can be 
neglected, hence the name \textit{quasistatic} ionization. 

This leads to a considerable simplification because one can work with a time-independent 
Hamiltonian. For the hydrogen atom, an estimate of the tunnel ionization rate was obtained 
already in 1928\cite{oppenheimer28} and later refined and extended.\cite{landau77,yamabe78,
ammosov86,tong02,batishchev10,tolstikhin14,yue17} The accurate evaluation of molecular 
quasistatic ionization rates has remained an active research field until today and is of relevance 
to many experiments where strong fields are applied. Especially the modeling of the angular 
dependence of the ionization rate is a challenge. This is caused by the exponential dependence 
of the ionization rate on the binding energy; in a polyatomic molecule the contribution of individual 
channels to the overall ionization rate varies greatly depending on orientation.\cite{jagau18,
hernandez19,hernandez20} 


\subsection{Shape and Feshbach resonances} \label{sec:restype}
\begin{figure} \centering
\includegraphics[scale=0.49]{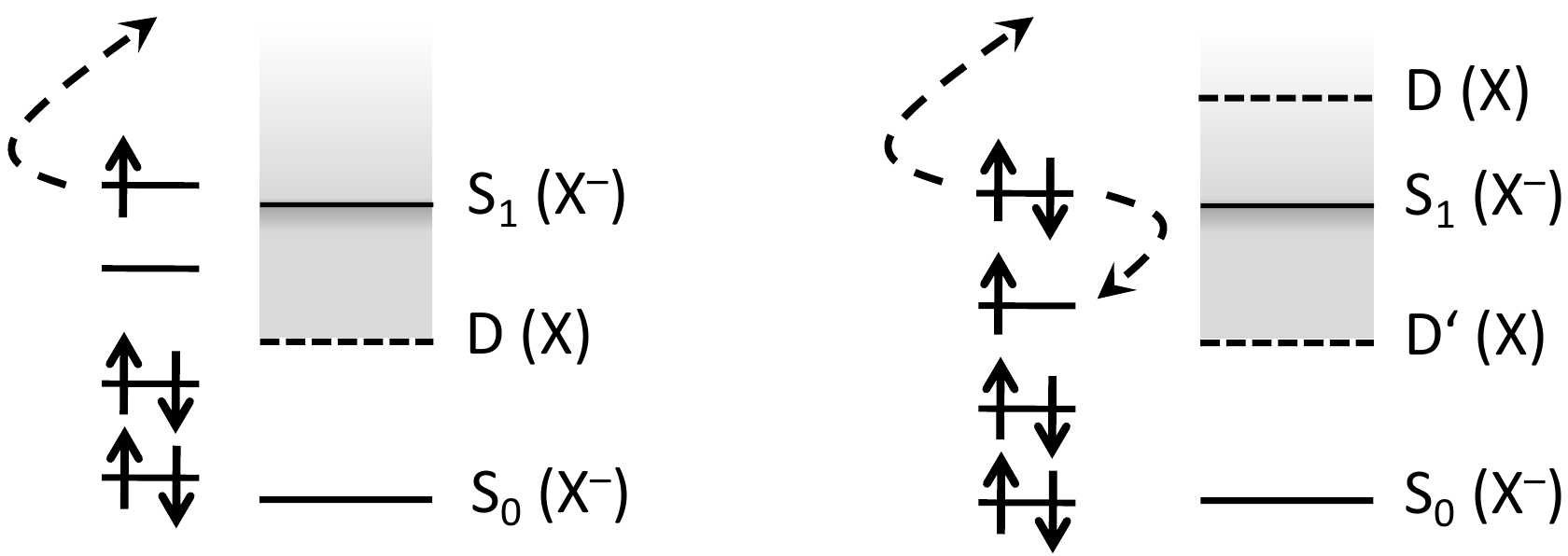}
\caption{Electronic structure of shape (left) and Feshbach (right) resonances. If the metastable 
S$_1$ state of the anion X$^-$ lies above its decay channel D, the decay is a one-electron 
process. If it lies below state D of the neutral molecule and at the same time above another 
decay channel D', the decay is a two-electron process.}
\label{fig:restype}
\end{figure}

One can group metastable electronic states into shape resonances, which decay by a one-electron 
process, and Feshbach resonances, which decay by a two-electron process.\cite{nhqmbook,
jagau17} The distinction is illustrated by Fig. \ref{fig:restype} for autodetaching anions. Notably, 
polyatomic molecules usually feature resonances of both types. Examples of shape resonances 
include \vspace{-0.1cm}
\begin{itemize}
\item temporary radical anions formed by electron attachment to neutral ground states,\cite{
simons08,simons11,herbert15} \vspace{-0.2cm}
\item low-lying excited states of closed-shell anions,\cite{simons08,simons11,herbert15} \vspace{-0.2cm}
\item most metastable dianions and more highly charged anions,\cite{dreuw02} \vspace{-0.2cm}
\item some superexcited states of neutral molecules,\cite{platzman62,hatano03} \vspace{-0.2cm}
\item core-excited states above the respective core-ionization threshold,\cite{piancastelli87,
kempgens99} \vspace{-0.2cm}
\item Stark resonances formed in static or dynamic electric fields.\cite{reinhardt76,scrinzi99,
scrinzi05,gallmann12,majety15b} 
\end{itemize}
Examples of Feshbach resonances include \vspace{-0.1cm}
\begin{itemize}
\item core-ionized states that undergo Auger decay,\cite{meitner22,auger23,xraybook} \vspace{-0.2cm}
\item core-excited states that undergo resonant Auger decay,\cite{brown80,armen00} \vspace{-0.2cm}
\item related species involving more than one molecule that decay through intermolecular 
Coulombic decay,\cite{cederbaum97,jahnke20} \vspace{-0.2cm}
\item anionic states formed by electron attachment to Rydberg states,\cite{schulz73,
ibanescu07,ibanescu08} \vspace{-0.2cm}
\item superexcited Rydberg states,\cite{platzman62,klinker18,plunkett19,rabadan21} \vspace{-0.2cm}
\item states of high spin multiplicity that decay only by spin-orbit coupling.\cite{sommerfeld98b,
dreuw99,dreuw99b,dreuw99c}
\end{itemize}
 
Shape resonances can be easily understood in real coordinate space: The shape of the effective 
potential is responsible for the metastable nature of the resonance, meaning the electron is 
trapped behind a potential wall but it can leave the system by tunneling.\cite{nhqmbook} In 
the case of radical anions this potential wall is given by the sum of the centrifugal potential 
and the molecular Coulomb potential,\cite{simons08,herbert15} whereas in dianions a repulsive 
Coulomb barrier is formed by the sum of short-range attractive valence interactions and 
long-range repulsion between every electron and the anionic rest.\cite{simons08,
herbert15} For Stark resonances, the potential wall is formed through the combination of the 
molecular potential and the external field.\cite{scrinzi05} 

For Feshbach resonances, decay by a one-electron process is energetically impossible. The 
metastability only originates from the coupling to another decay channel and 
involves more than one degree of freedom. Feshbach resonances are thus more naturally 
described in the space of electronic configurations.\cite{jagau17,nhqmbook} They can be 
viewed as superposition of two wave functions one of which is a bound states while the 
other one has continuum character. This also forms the motivation for the theory by Fano 
and Feshbach\cite{fano61,feshbach62} where one aims to decouple a 
Feshbach resonance from the continuum so that bound-state methods become amenable. 

Owing to the stronger coupling to the continuum, shape resonances are, in general, shorter 
lived than Feshbach resonances.\cite{nhqmbook,jagau17} However, this is not always the 
case because other factors influence the lifetime as well. Also, the character of a molecular 
resonance can vary across the potential surface as decay channels open or close. 


\section{Complex-variable techniques} \label{sec:cv}
\subsection{General theory} \label{sec:cvgen}
Since electronic resonances belong to the continuum, they cannot be described as stationary 
solutions of the time-independent Schr\"odinger equation in Hermitian quantum mechanics. 
Complex-variable techniques based on non-Hermitian quantum mechanics\cite{jagau17,
reinhardt82,nhqmbook} offer the possibility to describe electronic resonances in terms of 
discrete eigenstates. On a qualitative level, it is easy to see that the so-called Siegert 
representation\cite{siegert39} 
\begin{equation} \label{eq:sieg1}
E = E_R - i \, \Gamma/2
\end{equation}
describes a state with finite lifetime. The imaginary part of the energy leads to decay as the 
time evolution of the wave function shows: 
\begin{equation}  \label{eq:sieg2}
\Psi_\text{res} (t) = \exp [-i \, E \, t] \cdot \Psi_\text{res} (0) = \exp [- i\, E_R \, t ] \cdot 
\exp [- \Gamma \, t/2] \cdot \Psi_\text{res} (0)~.
\end{equation}
The probability density thus evolves in time as $-\Gamma \, t$ and the 
resonance width $\Gamma$ is connected to the characteristic lifetime $\tau$ through 
$\Gamma = 1/\tau$. Although there remain fundamental issues with non-Hermitian 
quantum mechanics that are yet to be solved, for example, the formulation 
of closure relations for non-Hermitian Hamiltonians,\cite{nhqmbook} Eq. \eqref{eq:sieg1} 
offers conceptual simplicity and emphasizes the similarity between resonance and 
bound-state wave functions. This equation forms the basis for extending quantum 
chemistry of bound states to electronic resonances. 

\begin{figure} \centering
\includegraphics[scale=0.60]{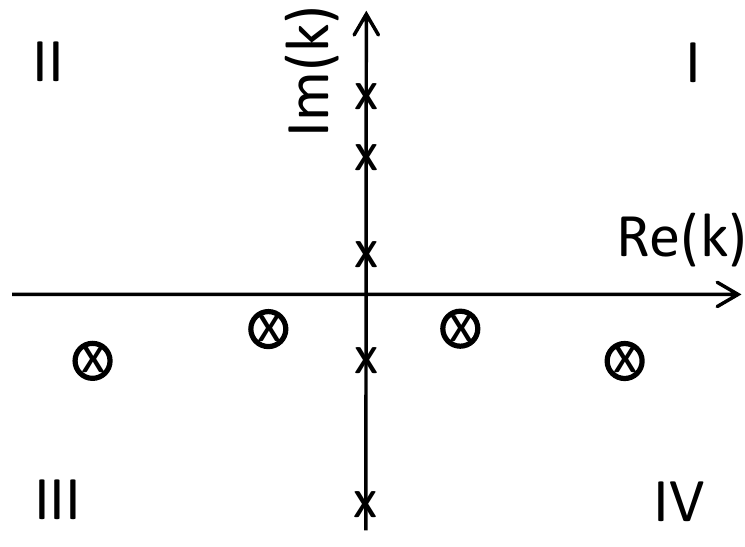}
\caption{Schematic representation of the complex momentum plane. Poles of S on the positive 
Im(k)-axis correspond to bound states ($\times$) whereas resonances ($\otimes$) lie in the 
fourth quadrant. Poles on the negative Im(k)-axis and in the third quadrant are antibound states 
($\times$) and capture resonances ($\otimes$), respectively. These states are 
in general beyond the reach of complex-variable methods and not discussed further here.
However, note that antibound states are also referred to as virtual states and related to 
s-wave scattering.\cite{taylor72}}
\label{fig:kplane}
\end{figure}

A more rigorous justification of Eq. \eqref{eq:sieg1} and the concept of complex-valued energies 
requires an analysis in terms of scattering theory, which is available, for example, in Refs. 
\citenum{nhqmbook,taylor72,domcke91}. The central quantity considered there is the scattering 
matrix $\mathbf{S}(k)$, which connects initial and final states of a system undergoing a scattering 
process, i.e., an electron-molecule collision in the case of electronic resonances. The poles of 
$\mathbf{S}(k)$ in the complex momentum plane, which represent the values of $k$ at which 
the amplitude of the incoming wave vanishes, can be associated with resonances and bound 
states as illustrated in Fig. \ref{fig:kplane}. 

Decaying resonances, meaning the poles of $\mathbf{S}(k)$ in the fourth quadrant 
in Fig. \ref{fig:kplane} are connected to peaks in the density of states in the continuum. 
The full width at half maximum $\Gamma$ of a peak centered at $E_R$ is given by Eq. 
\eqref{eq:sieg1} as can be derived by considering a closed contour integral in the lower half 
of the complex momentum plane\cite{nhqmbook,moiseyev98} 
\begin{equation}
N = \frac{1}{2\pi i} \oint_C \frac{\partial \text{ln} \mathbf{S}(k)}{\partial k} 
\mathrm{d}k 
\label{eq:contour}
\end{equation}
with $N$ as the number of poles in the lower half of the complex momentum 
plane, and applying the residue theorem to it. While a bound state with purely imaginary $k$ 
has an energy $E = k^2/2 \in \mathbb{R}^-$, a $k$-value in the fourth quadrant leads to an 
energy 
\begin{equation} \label{eq:sieg3}
E = 1/2 \cdot [\text{Re}^2(k) - \text{Im}^2(k) - 2\, i \, \text{Im}(k) \text{Re}(k)]~ \in \mathbb{C}~,
\end{equation}
where the imaginary part is strictly negative. The comparison of Eqs. \eqref{eq:sieg1} and 
\eqref{eq:sieg3} shows that the location of a resonance in the complex momentum plane is 
directly related to its position and width, i.e., $E_R$ and $\Gamma$. 

A resonance wave function corresponding to a pole of $\mathbf{S}(k)$ in 
the fourth quadrant in Fig. \ref{fig:kplane} behaves asymptotically, in the simplest case, like 
$\sim \exp [i \, \text{Re}(k) \, r] \; \exp [\text{Im}(k) \, r]$. This means that these wave functions 
diverge in space; they are thus outside the reach of quantum-chemical methods designed 
for $\mathcal{L}^2$ integrable states. Such outgoing boundary conditions 
as well as other boundary conditions used in scattering theory\cite{taylor72} are difficult 
to implement into quantum-chemistry software.\cite{nhqmbook,meyer82,
jagau17,masin20} An elegant solution is to regularize the diverging wave 
functions so that bound-state methods become applicable, which can be achieved by 
different techniques. Of interest to this article are complex scaling\cite{aguilar71,balslev71,
simon72,mccurdy78} and complex absorbing potentials,\cite{jolicard85,riss93} which are 
discussed in Secs. \ref{sec:cs} to \ref{sec:capbas}. Both approaches lead to a non-Hermitian 
Hamiltonian with complex eigenvalues that can be interpreted according to Eq. \eqref{eq:sieg1} 
and with corresponding eigenfunctions that are $\mathcal{L}^2$ integrable. 


Non-Hermitian Hamiltonians have, in general, different left and right eigenvectors.\cite{nhqmbook} 
However, when working with complex scaling or complex absorbing potentials, it is possible 
to choose identical left and right eigenvectors if the Hamiltonian is real-valued before the 
complex-variable treatment is applied. This implies that the matrix representation becomes 
complex symmetric. As a consequence, the usual scalar product needs to be replaced by the 
so-called $c$-product\cite{nhqmbook,reinhardt82,moiseyev78,moiseyev98}
\begin{equation} \label{eq:cprod}
\langle \Psi_i (r) | \Psi_j (r) \rangle = \int dr \, \Psi_i (r) \, \Psi_j (r)
\end{equation}
where the state on the left is not complex conjugated. The $c$-product is sometimes denoted 
by round brackets $(\dots | \dots)$ instead of angle brackets $\langle \dots | \dots \rangle$ but 
this practice is not followed here to avoid confusion with Mulliken and Dirac notation for 
electron-repulsion integrals. Instead, angle brackets are kept and the use of the $c$-product 
is always implied when discussing complex-variable methods. 

A further consequence of non-Hermiticity is the complex-variational principle,\cite{braendas77,
moiseyev78,moiseyev82} 
\begin{equation} \label{eq:cvar}
\tilde{E} = \langle \tilde{\Psi} | H | \tilde{\Psi} \rangle / \langle \tilde{\Psi} | \tilde{\Psi} \rangle ~ \in 
\mathbb{C}
\end{equation}
that holds for any $c$-normalizable trial wave function $\tilde{\Psi}$. Eq. \eqref{eq:cvar} replaces 
the usual variational principle for all complex-variable methods alike and is a stationarity principle 
for the complex energy rather than an upper or lower bound for its real or imaginary part. 


\subsection{Formal aspects of complex scaling} \label{sec:cs}
\begin{figure} \centering 
\includegraphics[scale=0.47]{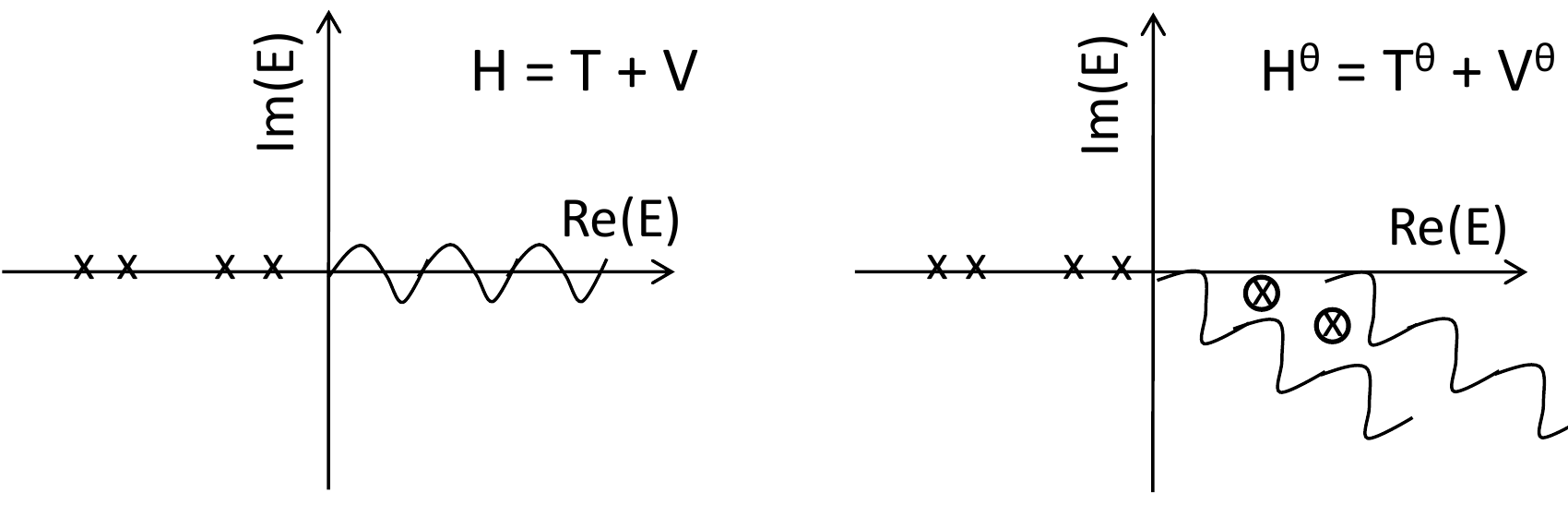} \hspace{1cm} \\[0.3cm]
\includegraphics[scale=0.47]{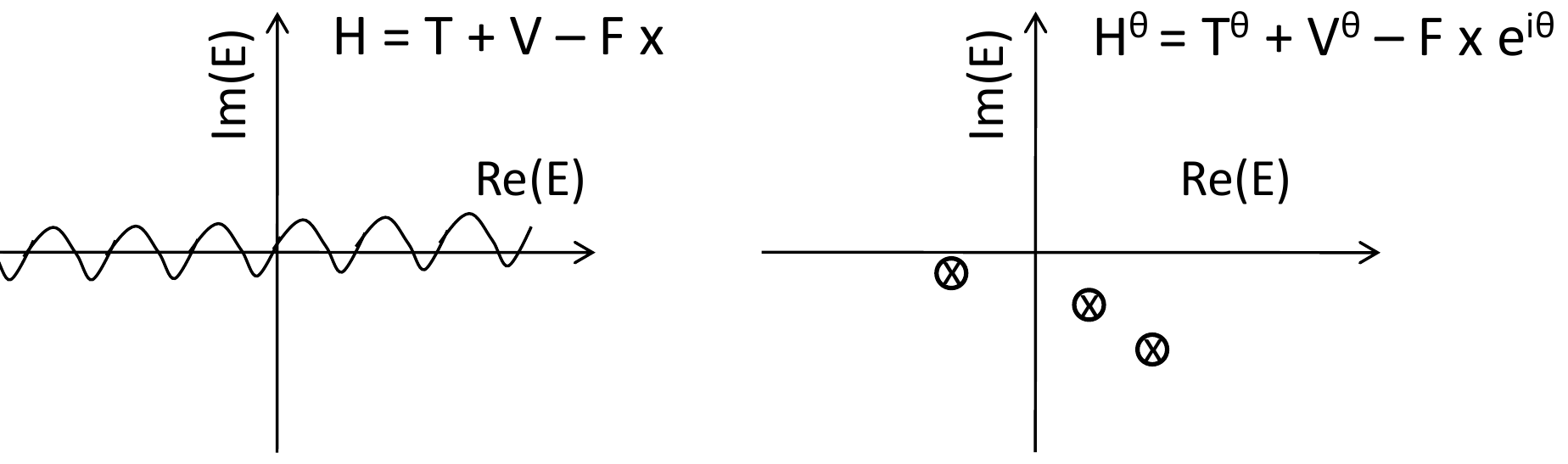}
\caption{Upper panels: Eigenvalue spectra of a Hamiltonian $H=T+V$ describing an autoionizing 
resonance and of its complex-scaled counterpart $H^\theta$. Lower panels: Eigenvalue spectra of 
a Hamiltonian $H=T+V-Fx$ describing ionization in a static electric field and of its complex-scaled 
counterpart $H^\theta$. Continua are denoted by $\sim\!\!\sim$, bound states by $\times$, and 
resonances by $\otimes$. Note that the potential $V-Fx$ does not support any bound state.}
\label{fig:cs}
\end{figure}

Complex scaling\cite{nhqmbook,reinhardt82,aguilar71,balslev71,simon72,reed82,loewdin88,
loewdin89,moiseyev98} is a mathematically rigorous technique to make diverging resonance 
wave functions $\mathcal{L}^2$ integrable. It represents a dilation transformation 
and relies on analytic continuation of the Hamiltonian to the complex plane. 
Analytic continuation is a mathematical technique to extend the domain over which an 
analytic function is defined. Upon scaling all coordinates in a Hamiltonian $H=T+V$ with 
$T$ as kinetic energy and $V$ as compact potential\cite{reed82} by a complex number 
$e^{i\theta}$, $\theta \in \mathbb{R}$, $\theta < \pi/4$, a non-Hermitian operator $H^\theta$ 
is obtained with the spectral properties illustrated in the upper panels of Fig. \ref{fig:cs}: 
\vspace{-0.1cm}
\begin{itemize}
\item All discrete eigenvalues of $H$, that is, bound-state energies, are also eigenvalues of 
$H^\theta$. \vspace{-0.2cm}
\item The continuous spectrum of $H^\theta$ is $\bigcup_{E_t} + a e^{-2i\theta}$, $a \in 
\mathbb{R}^+$ where $E_t$ are the thresholds of $H$, that is, in the context of electronic-structure 
theory the ionization energies if one deals with a neutral molecule and the detachment energies 
for an anion. Note that, for $H^\theta$, there are separate continua that correspond 
to the different thresholds and associated decay channels, whereas there is just one continuum 
for $H$. \vspace{-0.2cm}
\item $H^\theta$ may have discrete complex eigenvalues in the wedge formed by the continuous 
spectra of $H$ and $H^\theta$. These can be associated with the resonances.
\end{itemize}

A resonance wave function with the asymptotic behavior $\sim \exp[i  k  r]$, $k\in \mathbb{C}$ 
becomes $\mathcal{L}^2$ integrable upon complex scaling if $\theta > 1/2 \; \text{atan} [\Gamma 
/2 \, (E_R - E_t)]$. Above the same critical value, the resonance energies are independent 
of $\theta$. Since complex scaling relies on analytic continuation, the original theory\cite{
aguilar71,balslev71} is only applicable to Hamiltonians with dilation-analytic potentials. 
Such potentials can be loosely defined as being analytic in the parameter 
of the dilation transformation, that is, $e^{i\theta}$ in the case of complex scaling. A rigorous 
definition of dilation analyticity is given in Refs. \citenum{aguilar71} and \citenum{reed82}.

Whereas it has been established that the Coulomb potential is dilation analytic, an important 
case of a non-analytic potential is $V(x) \sim x$,\cite{reinhardt76,herbst78,herbst79,herbst81,
nicolaides92,scrinzi99} which describes a static electric field in $x$ direction. This implies that 
a Hamiltonian for an atom or molecule in a field has other spectral properties;\cite{herbst78,
herbst79,herbst81} these are illustrated in the lower panels of Fig. \ref{fig:cs}: \vspace{-0.1cm}
\begin{itemize}
\item Neither $H$ nor $H^\theta$ have discrete real eigenvalues, that is, bound states. \vspace{-0.2cm}
\item The continuous spectrum of $H$ comprises the whole real axis, whereas the continuous 
spectrum of $H^\theta$ is empty. \vspace{-0.2cm}
\item $H^\theta$ may have discrete complex eigenvalues. These can be associated with Stark 
resonances. \vspace{-0.2cm}
\end{itemize}
Although $V(x) \sim x$ is not dilation analytic, complex scaling renders Stark resonances, which 
asympotically behave as Airy functions, $\mathcal{L}^2$ integrable. 


\subsection{Complex scaling in the context of molecular electronic-structure theory} \label{sec:cs2}
While implementations of complex scaling in which the wave function is represented on a 
numerical grid preserve many of the appealing formal properties of the exact theory,\cite{
mccurdy91,scrinzi93,rescigno00,rescigno05} the representation in a basis set of atom-centered 
Gaussian functions suffers from several problems:\cite{bravaya13} The discrete eigenvalues 
of $H^\theta$ become $\theta$-dependent, whereas only the rotated continua depend on 
$\theta$ in the exact theory. This poses a problem for the computation of interstate properties 
as two states can depend on $\theta$ in a different way. A further problem arises for Stark 
resonances because the continuous spectrum is empty as illustrated in the lower panels 
of Fig. \ref{fig:cs}. The projection of $H^\theta$ onto a finite basis set therefore gives rise to 
additional unphysical eigenvalues that do not correspond to either bound, resonance, or 
continuum states.\cite{jagau16} 

The dependence of the bound-state and resonance energies on $\theta$ can be traced back 
to an oscillatory behavior $\sim \exp [-i \, Z \, r \sin\theta]$ of the respective wave functions 
that is introduced by complex scaling.\cite{chuljian83} Importantly, this dependence of the 
wave functions on $\theta$ is also present in exact theory, but since it is difficult to represent 
in Gaussian bases, it gives rise to an artificial $\theta$-dependence of the energies in this 
case. The magnitude of the perturbation is best quantified by analyzing bound states for 
which $\text{Im}(E)$ is strictly zero in exact theory but of the order of $10^{-4} - 10^{-3}$ 
a.u. in typical calculations in Gaussian basis sets.\cite{bravaya13,jagau16,matz21}

$\theta$-dependence entails the need to find an optimal value, $\theta_\text{opt}$, which 
is usually done using the criterion\cite{braendas77,moiseyev78,moiseyev98}
\begin{equation} \label{eq:thopt}
\text{min} \; [dE/d\theta]
\end{equation}
because this derivative vanishes in exact theory. Eq. \eqref{eq:thopt} works equally well for 
temporary anions, core-vacant states, as well as Stark resonances. A step size of $1^\circ$ 
for $\theta$ is usually adequate to evaluate Eq. \eqref{eq:thopt} to sufficient 
accuracy. In addition, $\theta$ is confined to the interval $0 < \theta < \pi/4$ by theory and 
varies little among resonances with similar electronic structure and among different 
electronic-structure methods. A more pronounced dependence on the basis set is, however, 
sometimes observed. In effect, this means that ca. 10--15 calculations need to be performed 
to determine $\theta_\text{opt}$ if no information can be inferred from preceding investigations. 
This is the main reason for the increased computational cost of complex-scaled calculations. 
Further reasons are the use of complex algebra and the need to include 
additional diffuse functions in the basis set; standard bases typically yield bad results.\cite{
bravaya13,jagau16,jagau17,matz21} While there are basis sets that have been designed 
specifically for resonances,\cite{venkatnathan00,zdanska05,kapralova13} the addition 
of a few shells of even-tempered diffuse functions to a standard basis usually works as 
well.\cite{bravaya13,jagau17} 


\subsection{Complex basis functions} \label{sec:cbf}
Complex scaling in its original formulation is limited to atomic resonances, because the 
electron-nuclear attraction is not dilation analytic within the Born-Oppenheimer (BO) 
approximation.\cite{nhqmbook,jagau17} However, because only the asymptotic behavior of the 
resonance wave function matters for the regularization, the non-analyticity can be circumvented 
by so-called exterior complex scaling (ECS),\cite{simon79,moiseyev88,rom90} where only the 
outer regions of space, where the electron-nuclear attraction vanishes, are complex scaled. ECS 
has been implemented for a number of time-dependent approaches in which the wave function 
is represented on a numerical grid.\cite{mccurdy91,scrinzi93,rescigno00,mccurdy04,rescigno05,
scrinzi10,tao09,yip08,yip14,majety15,orimo18} Such approaches hold a lot of promise but the 
realization of ECS in basis set of atom-centered Gaussians is difficult because most techniques 
for AO integral evaluation are not applicable to the ECS Hamiltonian. 

To implement ECS in the context of Gaussian basis sets, one can exploit the identity
\begin{equation} \label{eq:cbf1}
E = \frac{\langle \Psi(r) | H^\theta(r \, e^{i\theta}) | \Psi(r) \rangle}{\langle \Psi(r) | \Psi(r) \rangle} = 
\frac{\langle \Psi(r \, e^{-i \theta}) | H(r) | \Psi(r \, e^{-i \theta}) \rangle}{\langle \Psi(r \, e^{-i \theta}) |
\Psi(r \, e^{-i \theta}) \rangle}~,
\end{equation}
that is, scaling the coordinates of the Hamiltonian as $r \to r e^{i \theta}$ is equivalent to scaling 
the basis in which the Hamiltonian is represented as $r \to r e^{-i \theta}$.\cite{mccurdy78,
moiseyev79} The right-hand side of Eq. \eqref{eq:cbf1} is equivalent to scaling the exponents 
of Gaussian basis functions by $e^{- 2 i \theta}$.\cite{mccurdy78} If one does 
not apply this procedure to all basis functions but rather adds to a given basis set a number of 
extra functions with complex-scaled exponents,
\begin{equation} \label{eq:cbf2}
\chi_\mu(r, A) = N_\mu(\theta) \, S_\mu(r_A) \, \text{exp} \big[ -\alpha_\mu\, e^{-2i\theta} \, r_A^2 \big]~,
\end{equation}
a basis-set representation of ECS is obtained. Alternatively, the functions from Eq. \eqref{eq:cbf2} 
can be interpreted as being centered not at the nuclei but in the complex plane. The so-defined 
complex basis function (CBF) method\cite{mccurdy78,rescigno80,mccurdy80,honigmann99,
honigmann06,honigmann10,white15,white15b,white17} is compatible with the BO approximation; 
its main technical challenge consists in the need to evaluate integrals over the non-standard 
basis functions from Eq. \eqref{eq:cbf2}.\cite{white15} Whereas most of the established 
techniques for AO integral evaluation apply to complex exponents, several changes are 
necessary for the computation of the Boys function, which forms the first step of the 
evaluation of the electron-repulsion integrals; an efficient implementation was introduced 
only recently.\cite{white15} 

Besides the applicability to molecules, CBF methods offer further advantages over complex 
scaling of the Hamiltonian: $\text{Im}(E)$ of bound states is typically much smaller (ca. $10^{-5}$ 
a.u.) indicating a better representation of the complex-scaled wave function. Also, while an 
optimal scaling angle $\theta_\text{opt}$ needs to be determined using Eq. \eqref{eq:thopt}, the 
values of $dE/d\theta$ are significantly smaller than when the Hamiltonian is complex-scaled. 
The requirements of CBF calculations towards the basis set are, in general, somewhat less 
arduous than those of complex scaling, although it is necessary to augment standard bases 
by extra functions. The details depend on the type of resonance and the energy released in 
the decay process.\cite{white15,white15b,white17,jagau18,matz21} 


\subsection{Formal aspects of complex absorbing potentials} \label{sec:cap}
Complex absorbing potentials (CAPs) have been used for long in quantum dynamics 
as a means to prevent artificial reflection of a wave packet at the edge 
of a grid.\cite{muga04} By accident, it was discovered that CAPs are also useful in static 
electronic-structure calculations.\cite{jolicard85} By absorbing the diverging tail of the 
wave function, a CAP enables an $\mathcal{L}^2$ treatment of resonances similar to 
complex-scaled approaches. More thorough analysis showed that CAPs are, under 
certain conditions, mathematically equivalent to complex scaling.\cite{riss93,riss95,riss98,
moiseyev98b} If this is the case, exact resonance positions and widths can 
be recovered from CAP calculations. Still, in practice, CAP methods feature more heuristic 
aspects than complex-scaled methods as will be detailed in this section and the subsequent 
one. At the same time, CAPs are easier to integrate into molecular electronic-structure 
theory.\cite{jagau17,zuev14} In particular, no difficulties arise about the combination with 
the BO approximation so that CAP methods can be readily applied to molecules. 

The CAP-augmented Hamiltonian is given as 
\begin{equation} \label{eq:cap1}
H^\eta = H - i \; \eta \; W
\end{equation}
with $\eta \in \mathbb{R}^+$ as strength parameter. Different functional forms have been 
suggested for $W$,\cite{riss93,riss95,riss98,moiseyev98b,rom91,sajeev06,sommerfeld01,
ghosh12,zuev14,sommerfeld15,sommerfeld98} in the simplest case $W$ is chosen as a 
real-valued quadratic potential of the form 
\begin{equation} \label{eq:cap2}
W = \!\! \sum_{\alpha=x,y,z} \!\!W_\alpha\, , ~\; W_\alpha = \begin{cases} (|r_\alpha - o_\alpha| 
- r_\alpha^0)^2 &\text{if} ~ |r_\alpha - o_\alpha| > r_\alpha^0 \\ 0 &\text{if} ~ |r_\alpha 
- o_\alpha| < r_\alpha^0 \end{cases}~.
\end{equation}
Here, the vector $(r_x^0, r_y^0, r_z^0)$ defines the onset of the CAP and hence a cuboid 
box in which the CAP is not active and the vector $(o_x, o_y, o_z)$ is the origin of the CAP. 
$r_\alpha^0$ and $o_\alpha$ are heuristic parameters, protocols for their determination are 
discussed in Sec. \ref{sec:capbas}. 

A suitable starting point for the mathematical analysis of $H^\eta$ is a free particle in one 
dimension exposed to a CAP with onset $r_x^0 = 0$.\cite{riss93,santra06} In this case, $H$ 
comprises only kinetic energy and the operator
\begin{equation} \label{eq:cap3}
H^\eta = - \frac{1}{2} \frac{\partial}{\partial x^2} - i \, \eta \, x^2 
\end{equation}
describes a harmonic oscillator with frequency $\sqrt{\eta} \, (1-i) = \sqrt{2\eta} \, \text{exp} 
[-i \pi/4]$ on a complex-rotated length scale $(2 \, \eta)^{-1/4} \, \text{exp} [i \, \pi/8]$. The 
spectrum of eigenvalues is $E_n = \sqrt{\eta} (1-i) \, (n+1/2), \; n \in \mathbb{N}$, that is, it is 
rotated into the lower half of the complex energy plane by an angle of $2\theta=\text{exp} 
[-i\, \pi/4]$ mimicking complex scaling with $\theta = \pi/8$. At the same time, the spectrum 
of $H^\eta$ is purely discrete for finite $\eta$; the continuous spectrum of the complex-scaled 
Hamiltonian (see upper panels of Fig. \ref{fig:cs}) is recovered only in the limit $\eta \to 0^+$. 
By similar considerations it is possible to show that general monomial CAPs $W\sim r^n$ 
correspond to complex scaling with an angle $\theta = -\pi/(2+n)$.\cite{riss93} Inclusion of 
a compact potential in $H^\eta$ does not fundamentally change this analysis. 

CAPs of a more elaborate form than Eq. \eqref{eq:cap2} have been investigated as well.\cite{
riss95,riss98,moiseyev98,rom91,sajeev06} One guiding principle has been to design potentials 
that are equivalent to complex scaling not only in the limit $\eta \to 0^+$ but also for finite values 
of $\eta$. As a first step, one can choose $\eta$ to be energy-dependent because it can be 
shown that reflection and absorption properties depend on the energy as well. Applying the 
ansatz $\eta(E) = 2 \, a \, (E-V)$ to Eq. \eqref{eq:cap1} leads to a modified Schr\"odinger 
equation\cite{riss95} 
\begin{equation} \label{eq:tcap1}
T |\Psi(x) \rangle + V(x) \, (1 + 2i \, a \, W(x)) | \Psi(x) \rangle = E (1 + 2i \, a \, W(x) ) 
| \Psi(x) \rangle 
\end{equation}
which can be recast in the usual form by introducing the function $f = \sqrt{1 + 2i\, a \, W}$. 
This results in 
\begin{equation}  \label{eq:tcap2}
f^{-1} \, T \, f^{-1} \, | \widetilde{\Psi}(x) \rangle + V(x) | \widetilde{\Psi}(x) \rangle 
= E | \widetilde{\Psi}(x) \rangle
\end{equation}
with $| \widetilde{\Psi}(x) \rangle = f \, |\Psi(x) \rangle$ as transformed wave function. Eq. 
\eqref{eq:tcap2}, which features a modified kinetic energy, illustrates the connection between 
CAPs and ECS; it can be shown that the function $f$ defines a scaling contour in the complex 
plane. Since the theory of complex scaling places only few restrictions on the scaling contour,\cite{
nhqmbook,moiseyev98} that is, the path in the complex plane along which the 
coordinates are scaled, diverse functional forms of $W$ are possible. This connection between 
CAPs and ECS has triggered the development of the \textit{transformative} CAP\cite{riss95,
riss98} (TCAP) and of several \textit{reflection-free} CAPs\cite{riss95,riss98,moiseyev98b,
sajeev06} all of which are built around a transformed kinetic energy. The TCAP has been 
implemented for a few electronic-structure methods,\cite{sommerfeld98} which showed that, 
despite formal advantages over Eq. \eqref{eq:cap2}, the results still suffer from the same 
shortcomings when using atom-centered Gaussian basis sets. Most recent developments 
are thus built on the simpler functional form from Eq. \eqref{eq:cap2}. 


\subsection{Complex absorbing potentials in the context of molecular electronic-structure 
theory} \label{sec:capbas}
While the limit $\eta \to 0^+$ can be taken if the Schr\"odinger equation is solved exactly, 
this is not the case when $H^\eta$ is represented in a finite basis set and 
the Schr\"odinger equation is solved approximately.\cite{riss93} Instead, one has to use 
finite $\eta$ values and balance out two effects: The error caused by the finite CAP strength, 
which increases with $\eta$ and an additional error $\Delta E_\text{bas}(\eta)$ caused by 
the finite basis set, which decreases with $\eta$. The optimal value $\eta_\text{opt}$ is 
usually determined by a Taylor expansion of the resonance energy in $\eta$ 
\begin{equation} \label{eq:capbas}
E(\eta) = E_0 + a_1 \eta + a_2 \eta^2 + a_3 \eta^3 + \dots + \Delta E_\text{bas} (\eta) 
\end{equation}
and minimizing the linear term, which leads to the criterion\cite{riss93} 
\begin{equation} \label{eq:capopt}
\text{min} \, | \eta dE / d\eta |~.
\end{equation}
In analogy to complex-scaled methods, this entails the need to calculate trajectories $E(\eta)$. 
However, CAP theory does not place an upper limit on $\eta$, whereas the complex scaling 
angle is restricted to $0 < \theta < \pi/4$. In practice, $\eta_\text{opt}$ varies from $10^{-4}$ 
to $10^{-1}$ a.u. between different calculations depending on the completeness 
of the basis set and the choice of onset  $(r_x^0, r_y^0, r_z^0)$ and origin $(o_x, o_y, o_z)$ 
so that, typically, between 20 and 50 calculations need to be run to determine the optimal CAP 
strength.\cite{jagau17,zuev14,zuev14e} 

For temporary anions, a possible choice for the onset that avoids optimization on a case-by-case 
basis consists in $\sqrt{X^2}$, $\sqrt{Y^2}$, $\sqrt{Z^2}$ where the expectation values are 
computed for the neutral ground state.\cite{zuev14,zuev14e,jagau14b,jagau14e} This approach 
aims to minimize the perturbation of the neutral system and to apply the CAP 
only to the extra electron. It is, however, not straightforward to generalize the idea to other types 
of resonances beyond temporary anions. As an alternative, the definition of the CAP onset on the 
basis of Voronoi cells has been suggested.\cite{sommerfeld15,ehara16} This in advantageous for 
larger molecules of irregular shape where the definition of a CAP according to Eq. \eqref{eq:cap2} 
is questionable. In other cases, Voronoi CAPs and box-shaped CAPs deliver very similar results. 
The impact of the CAP origin has been investigated less thoroughly, it is usually chosen as the 
center of mass or the center of nuclear charges of a molecule.\cite{zuev14,benda17,benda18a} 

While the integrals $W_{\mu\nu} = \langle \chi_\mu | W | \ \chi_\nu \rangle$ of a box-shaped 
CAP defined according to Eq. \eqref{eq:cap2} over Gaussian functions can be calculated 
analytically,\cite{santra01} the corresponding integrals of a Voronoi CAP need to be evaluated 
numerically.\cite{sommerfeld15} However, the evaluation of integrals over arbitrary potentials 
is a standard task in density functional theory and adapting such functionalities to the evaluation 
of CAPs is straightforward.\cite{zuev14} In general, the evaluation of $W_{\mu\nu}$ does not 
drive the computational cost. 

In analogy to complex-scaled and CBF calculations, additional shells of diffuse functions 
need to be added to a basis set to represent the resonance wave function properly in a 
CAP calculation. The same basis sets usually work as well but it is sometimes possible to 
truncate the basis set based on physical considerations when using CAPs.\cite{jagau17,
zuev14} For example, temporary anions that arise from adding an electron to a $\pi^*$ 
orbital can often be described with a basis set that includes only additional $p$ functions. 
However, a general observation is that the use of too small basis sets in CAP calculations 
gives rise to additional spurious minima in Eq. \eqref{eq:capopt} that do not correspond to 
physical resonance states.\cite{jagau14b,zuev14,jagau17} 

Several ideas have been proposed to reduce the dependence of the resonance energy on 
the CAP parameters. One can, for example, compute the linear term in Eq. \eqref{eq:capbas} 
explicitly and subtract it from the energy.\cite{riss93,jagau14b} This leads to the first-order 
corrected energy 
\begin{equation} \label{eq:capcorr}
E^{(1)} (\eta) = E(\eta) - \eta \, dE/d\eta = E + i \, \eta \, \langle W \rangle~,
\end{equation}
where the correction term is given as the expectation value of the CAP operator. 
$\eta_\text{opt}$ can be determined for Eq. \eqref{eq:capcorr} by minimizing the 
next-higher order term in Eq. \eqref{eq:capbas}, which leads to the criterion $\text{min} 
|\eta \, dE^{(1)}/d\eta|$. However, the correction leads to an increased basis-set error 
$\Delta E_\text{bas}(\eta)$ so that it is not guaranteed that $E^{(1)}$ represents an 
improvement over the uncorrected energy $E$.\cite{riss93} In practice, this can be tested 
by comparing the values of $|\eta \, dE/d\eta|$ and $|\eta \, dE^{(1)}/d\eta|$, which shows 
that, in general, Eq. \eqref{eq:capcorr} does represent an improvement. The first-order 
corrected energy is less dependent on $\eta$ and the CAP onset than the uncorrected 
energy.\cite{jagau14b,jagau14e,jagau17} 

As an alternative to the Taylor expansion in Eq. \eqref{eq:capbas}, it has been suggested 
to express $E(\eta)$ in terms of Pad\'e approximants.\cite{lefebvre05,landau16,landau19} 
This allows to recover the non-analytic limit $\eta \to 0^+$ from a series of calculations 
with different $\eta$ values; the dependence of the energy on $\eta$ is, however, also 
with this approach not removed completely. 

Recently, the integration of CAPs into Feshbach-Fano theory has been suggested\cite{
kunitsa19} in order to define the projector needed there for separating bound and 
continuum parts of the wave function. In analogy to other approaches based on this 
theory, the computational cost is lower as compared to complex-variable methods 
because the Schr\"odinger equation is solved with a real-variable electronic-structure 
method and the CAP is added afterwards. However, only very few applications of this 
approach have been reported so far.\cite{kunitsa19}


\subsection{Practical aspects of complex-variable electronic-structure calculations} \label{sec:elst}
The choice of electronic-structure model is of central importance for the accurate 
description of molecular resonances. Complex scaling and complex absorbing potentials 
have been combined with a variety of methods; implementations of Hartree-Fock 
(HF),\cite{rescigno80,mccurdy80,white15b} density functional theory (DFT),\cite{
whitenack11,whitenack12,zhou12},multireference configuration interaction (MRCI),\cite{
sommerfeld98,sommerfeld01,honigmann06,honigmann10}  resolution-of-the-identity 
second-order M{\o}ller-Plesset perturbation theory (RI-MP2),\cite{hernandez19,
hernandez20} extended multiconfigurational quasidegenerate perturbation theory 
(XMCQDPT2),\cite{kunitsa17} algebraic diagrammatic construction (ADC),\cite{santra02,
feuerbacher03,belogolova21,dempwolff21} symmetry-adapted-cluster configuration 
interaction (SAC-CI),\cite{ehara12} multireference coupled-cluster (MRCC),\cite{sajeev05} 
and equation-of-motion (EOM)-CC methods\cite{ghosh12,bravaya13,zuev14,white17} 
have been reported. Since there exist different ways to set up a complex-variable calculation, 
the comparison of results obtained using different implementations is not straightforward. 
Moreover, many approaches are specifically designed for the treatment of a particular 
type of resonance, i.e., only suitable for temporary radical anions, core-vacant states, or 
molecules in static electric fields. This has led to a situation where many resonances have 
been investigated with only one or two computational approaches. Systematic benchmarks, 
where the same resonances are computed using the same basis sets and complex-variable 
techniques but different electronic-structure models are still largely missing. The same 
applies to the numerical comparison of CAPs and complex scaling calculations based 
on the same electronic-structure model.

A first requirement towards a complex-variable electronic-structure model is that the wave 
function comprises configurations that describe the decay of the state. This is usually easy 
to achieve for shape resonances but requires more consideration in the case of Feshbach 
resonances. For example, the configurations describing Auger decay of a core-ionized state 
are doubly excited with respect to the core-vacant HF determinant (see Fig. \ref{fig:auger}); 
as a consequence, the decay cannot be described with HF theory. 

A second requirement is to achieve a consistent description of the resonance itself and its 
decay channels.\cite{jagau14} Multistate methods such as EOM-CC,\cite{ccbook,krylov08,
sneskov12,bartlett12,emrich81,sekino84,stanton93,nooijen93,stanton94} CC2,\cite{
christiansen95,haettig00,haettig06} and ADC(n),\cite{dreuw15,schirmer82,trofimov95,
trofimov99} which deliver wave functions for states with $N$ and $N+1$ electrons within 
the same computation, have a built-in advantage because they place the ionization 
continua such that no ambiguity over the character of a state arises. In contrast, if one 
performs, for example, two separate HF calculations for a temporary radical anions and 
the corresponding neutral molecule, it can happen that the decay width of the anion is 
zero although its HF energy is higher than that of the neutral species. This is further 
discussed in the context of analytic-gradient theory in Sec. \ref{sec:grad}. 

\begin{figure} \centering
\includegraphics[scale=1.25]{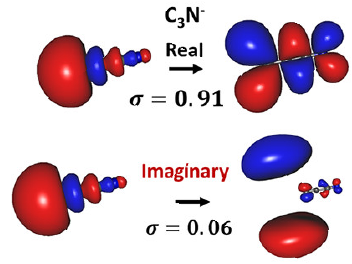}
\caption{Real and imaginary NTOs and their corresponding singular values 
$\sigma_K$ for the $^3\Pi$ state of C$_3$N$^-$ computed with CAP-EOM-EE-CCSD. 
Reproduced with permission from Ref. \citenum{skomorowski18}.}
\label{fig:orbs}
\end{figure}

Another advantage of multistate methods is the straightforward evaluation 
of transition moments as well as Dyson orbitals\cite{jagau16b} and natural transition orbitals 
(NTOs).\cite{skomorowski18} These quantities are equally relevant for resonances as for 
bound states;\cite{krylov20} for example, they enable the application of exciton 
theory to resonances, which helps explain spectroscopic signatures of resonances.\cite{
skomorowski18} As a consequence of Eq. \eqref{eq:cprod},the real and imaginary parts 
of orbitals need to be interpreted separately in the context of complex-variable methods. 
A representative example of an NTO pair is shown in Fig. \ref{fig:orbs}.

A further advantage of multistate methods is technical but crucial for some 
types of states: It is often possible to avoid solving the SCF equations for the resonance. 
Especially for temporary radical anions, it can be very difficult to ensure convergence to 
the resonance state instead of a pseudocontinuum solution. For other types of resonances, 
however, this is less problematic. For example, CBF-HF determinants for a core-vacant 
state or a molecule in a static electric field can be easily constructed using maximum-overlap 
techniques\cite{gilbert08} starting from a real-valued core-vacant HF determinant or a field-free 
HF determinant, respectively.\cite{jagau16,jagau18,matz21} 

It also needs to be mentioned here that the strengths and weaknesses that have been established 
for a particular electronic-structure method in the context of bound states, are still relevant for 
resonances. For example, many resonances have open-shell character, a spin-complete 
description as afforded by multistate methods such as EOM-CC and ADC is advantageous. 

From these considerations, a number of computational approaches arise that combine several 
advantages for different types of resonances: For temporary radical anions, EOM-EA-CC within 
the singles and doubles approximation (EOM-EA-CCSD) based on a CCSD 
reference for the neutral closed-shell state is well suited or, alternatively, the EA variant of ADC(2) 
or ADC(3) starting from an MP2 or MP3 reference. Likewise, excited states of closed-shell anions, 
superexcited Rydberg states, and core-excited states are best described with the EE variants 
of the same methods, while the IP variants are suitable for core-ionized states. 

For other types of states, a fully satisfactory description is more difficult to achieve: For example, 
in the case of closed-shell dianions, one can construct a CBF-HF or CAP-HF determinant for 
the resonance, treat correlation by means of CCSD, and describe the detachment channels as 
EOM-IP-CCSD states.\cite{gulania19} This delivers a balanced and spin-complete description 
of all relevant states but the bound monoanionic states, which are described using orbitals of 
the resonance state, will likely have substantial non-zero imaginary energies. The alternative 
approach, where one starts from an open-shell HF determinant for one of the monoanionic 
decay channels, avoids the latter problem but delivers a description that is neither balanced 
nor spin-complete.  A similar problem exists for Stark resonances where the Hamiltonian does 
not feature any bound states; it is thus inevitable to solve the SCF equations directly for the 
resonance.\cite{jagau16,jagau18}

In post-HF CAP methods, a further degree of freedom arises because the CAP does not 
need to be added to the Hamiltonian from the outset. For example, in a CAP-EOM-CC 
treatment, the CAP can be introduced already in the HF calculation,\cite{zuev14,zuev14e} 
or at the CC step, or even only at the EOM-CC step.\cite{ghosh12} No such choice exists 
for CBF methods and one needs to work with a complex-valued wave function throughout. 
In the full CI limit, all three variants of CAP-EOM-CC deliver identical results but this is not 
the case for truncated CC methods. The distinction between relaxed and unrelaxed (EOM-) 
CC molecular properties\cite{gauss00} is related to this subject and, similar to what applies 
there, the numerical differences between the three CAP-EOM-CC variants are usually small.

Also, there is no clear formal advantage of one scheme over the other: If the CAP is active 
already at the HF level, the form of the CC and EOM-CC equations does not change as 
compared to the real-valued formalism. Also, the size-extensivity of truncated CC methods 
is preserved and it is straightforward to work out analytic-derivative theory (see Sec. 
\ref{sec:grad}). The main advantage of the alternative variant in which the CAP-EOM-CC 
Hamiltonian is built from a real-valued CC wave function is its reduced computational cost. 
Only the EOM-CC equations have to be solved for multiple values of the CAP strength to 
evaluate Eq. \eqref{eq:capopt}, the CC equations for the reference state need to be solved 
only once. However, these methods are not size-intensive and it is necessary to include 
additional terms in the EOM-CC equations because the cluster operator does not fulfill the 
CC equations at $\eta\neq 0$. 

As a further idea in the context of temporary radical anions, it has been suggested to 
project the CAP on the virtual orbital space.\cite{santra99} The rationale is to minimize 
perturbation of the occupied orbital space and to apply the CAP only to the extra electron. 
This has been realized for CI and EOM-CC methods;\cite{santra99,ghosh12,jagau17} 
numerical evidence suggests that the impact on the results is small. 

Finally, it should be mentioned that there is some ambiguity in the evaluation of Eqs. 
\eqref{eq:thopt} and \eqref{eq:capopt} as one can search for minima in the total resonance 
energy or in the energy difference with respect to some bound parent state.\cite{bravaya13,
white17} If the Schr\"odinger equation was solved exactly, this would not matter because 
bound-state energies do not depend on the complex scaling angle or the CAP strength 
in that case. However, it can matter for approximate solutions, in particular for calculations 
with a complex-scaled Hamiltonian where bound states acquire substantial imaginary 
energies (see Sec. \ref{sec:cs2}). For CBF calculations, the impact is substantially smaller. 
Numerical experience shows that Eq. \eqref{eq:thopt} is best applied to energy differences 
here.\cite{white17,jagau18,matz21} For CAP calculations, the differences are usually 
negligible. 


\section{Recent methodological developments} \label{sec:dev}

\subsection{Complex-valued potential energy surfaces and analytic gradient theory} \label{sec:grad}
The concept of potential energy surfaces (PES) is a cornerstone of the quantum chemistry of bound 
states. By virtue of the BO approximation, the electronic energy is obtained as a function of the 
nuclear coordinates by solving the electronic Schr\"odinger equation at fixed nuclear positions. 
The nuclear dynamics are then described in terms of these PES.

Consideration of nuclear motion is equally important for molecular electronic resonances because it 
happens in many cases on the same timescale as electronic decay. As an example, 
consider DEA to a closed-shell molecule. The cross section for this process is determined by the 
interplay of nuclear motion and electronic decay.\cite{fabrikant17} Moreover, there are important 
cases of DEA where two electronic resonances are coupled through nuclear motion\cite{estrada86,
feuerbacher04}. Aside from DEA, there are anions whose autodetachment is entirely due to nuclear 
motion,\cite{simons81,simons99,simons02} for example NH$^-$\cite{chalasinski88} and enolates,\cite{
oneal88} as well as other anions that are adiabatically bound only when zero-point vibrational energies 
are taken into account, for example benzonitrile.\cite{gulania20} Vibrational effects hence play a 
decisive role for the spectroscopy of temporary anions,\cite{simons11} but also for other types of 
resonances such as core-vacant states.\cite{norman18}

\begin{figure} \centering
\includegraphics[scale=0.38]{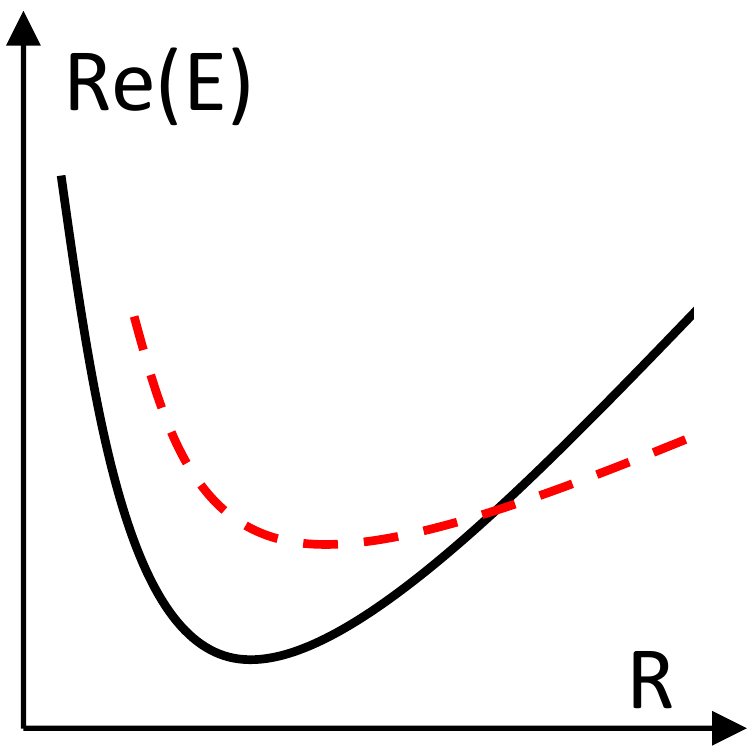} \hspace{0.8cm}
\includegraphics[scale=0.38]{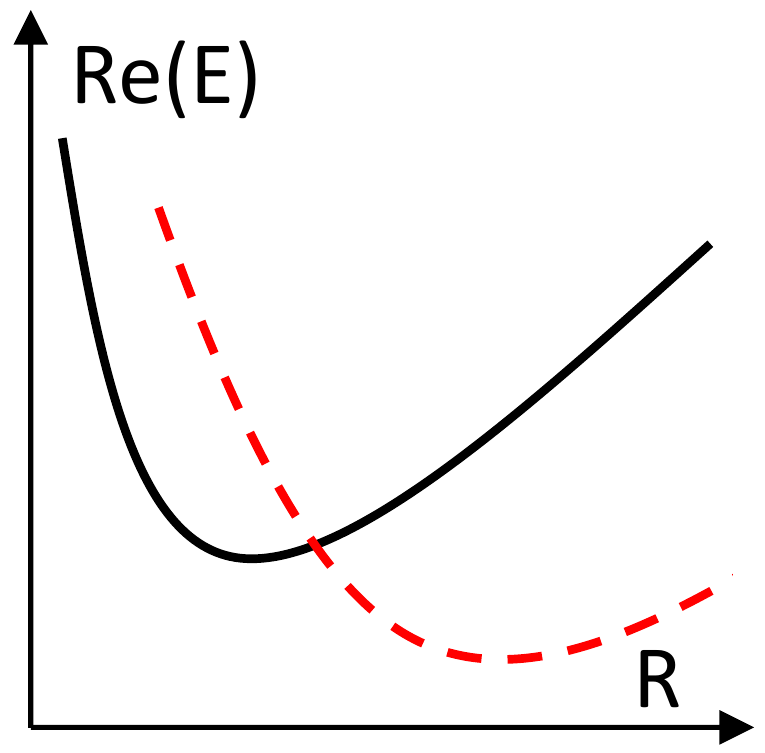} \\
\includegraphics[scale=0.38]{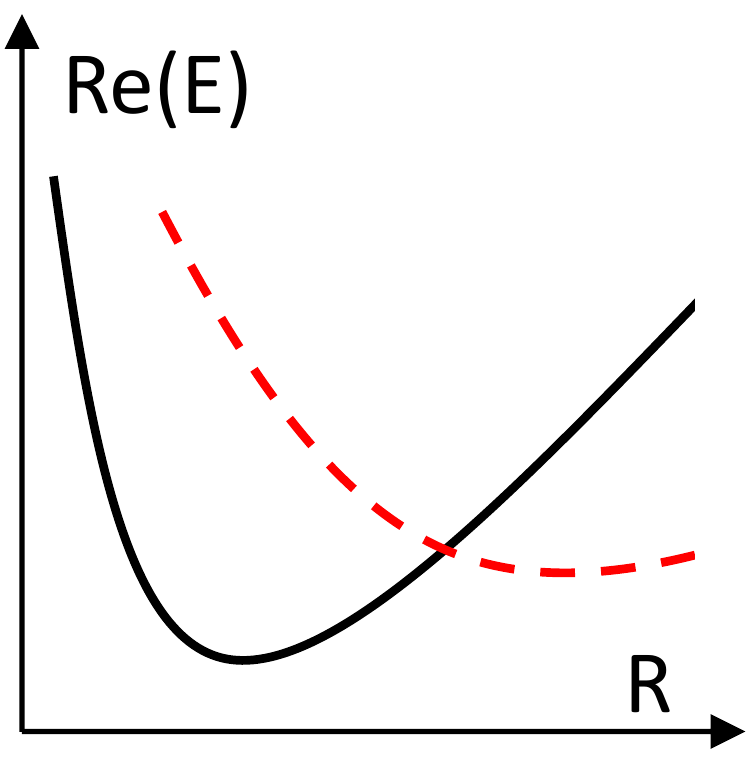} \hspace{0.8cm}
\includegraphics[scale=0.38]{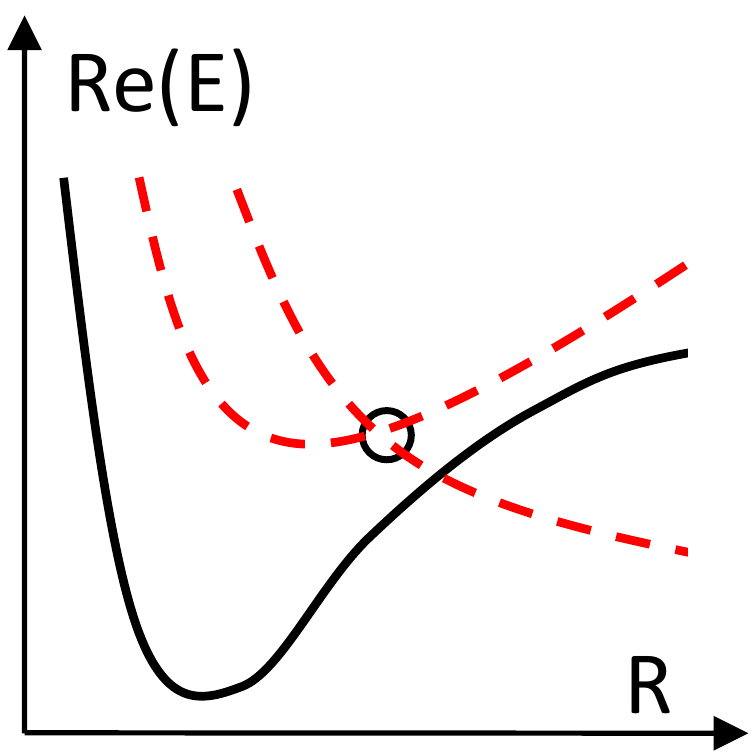}
\caption{Exemplary shapes of potential energy surfaces of temporary anions in one dimension. 
Dashed red lines denote anionic states, solid black curves denote neutral states. Upper left: The 
anion is unbound but vertically stable at stretched bond lengths. (Example: N$_2^-$) Lower left: 
The anion is vertically stable at its own equilibrium structure but adiabatically 
unbound (Example: CO$_2^-$) Upper right: The anion is adiabatically bound but 
vertically unstable near the equilibrium structure of the parent neutral state. (Example: F$_2^-$) 
Lower right: Dissociative electron attachment is possible and proceeds through coupling of two 
resonances. The exceptional point is marked by a circle. (Example: HCN$^-$)}
\label{fig:cpes}
\end{figure}

Using the Siegert representation, Eq. \eqref{eq:sieg1}, the concept of PES can be readily generalized 
to resonances: The real part of the PES is interpreted in analogy to bound states, 
whereas the imaginary part yields the decay width as a function of the molecular structure.\cite{
moiseyev17} Complex-valued resonance PES can be as diverse as those of bound states but the 
former are in general much less well characterized than the latter. Fig. \ref{fig:cpes} illustrates several 
typical PES shapes for temporary radical anions. These resonances often become stable towards 
electron loss through structural changes, typically through bond stretching.\cite{jagau17} However, 
such stabilization does not pertain to most other types of resonances: A molecule in a static electric 
field does not have any bound states so that the electronic energy remains complex-valued at all 
nuclear configurations. The same applies to core-vacant states, which also do not become bound 
through structural rearrangement. 

Plots such as those in Fig. \ref{fig:cpes} cannot represent the full dimensionality 
of the PES for systems with more than two atoms; they can be exact only for diatomic molecules. 
An $N$-atomic molecule has $M=3N-6$ internal nuclear degrees of freedom ($3N-5$ for linear 
molecules); the PES is thus a high-dimensional object. In these cases, plots such as those in 
Fig. \ref{fig:cpes} only represent cuts through the PES and do not capture the full dimensionality. 
Although low-dimensional models can afford meaningful insights, it is desirable to treat all $M$ 
degrees of freedom on an equal footing as is routinely possible for bound states. 

Bound-state PES are commonly characterized in terms of special points such as equilibrium 
structures, transition structures, conical intersection seams, and minimum 
energy crossing points (MECPs).\cite{matsika21,koeppel84,domcke11} Analogous special 
points on complex-valued PES are of interest for resonances as well. Equilibrium structures 
are characterized by $dE_R/d\mathbf{R} = 0$ and positive real parts of all eigenvalues of the 
Hessian matrix; $\Gamma$ is not relevant here. 

Crossing seams between a temporary anion and its parent neutral state mark 
the region where the anion becomes stable towards electron loss. Since the two states involved 
in such a crossing have a different number of electrons, the Hamiltonian does not couple 
them and the crossing seam has the dimension $M-1$. If $E_R$ and $E_0$ are obtained 
as eigenvalues of the same Hamiltonian, as is possible, for example, with EOM-CC or ADC 
methods, $\Gamma$ in principle becomes zero exactly at this crossing seam,\cite{jagau14} 
which is not the case if $E_R$ and $E_0$ are computed independent of each other. However, 
since very small decay widths are difficult to represent with CAP methods, deviations are 
observed in practice also for multistate methods.\cite{benda18b}

The crossing seam between a temporary anion and its parent state can be related to the 
stability of a molecule towards low-energy electrons and the efficiency of DEA 
in a similar way in which the location and topology of a conical intersection 
explain photostability. Along the seam, the MECP is of particular interest, which motivated 
the development of an algorithm for locating it.\cite{benda18b} In a straightforward extension 
of a similar algorithm for bound states,\cite{bearpark94} this can be done 
using the condition\cite{benda18b}
\begin{equation} \label{eq:mecp}
d/d\mathbf{R} \Big[E_R - E_0 \Big]^2 = 0
\end{equation}
and at the same time minimizing the energy of one of the states in the space orthogonal to 
the crossing seam. 


Also of interest are crossing seams between two resonances. The intersection 
space where the real and imaginary parts of the energies are degenerate is termed exceptional 
point (EP).\cite{kato66,heiss12} As for conical intersections, the dimension of 
an EP seam is $M-2$,\cite{kato66,benda18c} which implies that diatomic molecules cannot 
feature EPs. More general intersection spaces, where only the real \textit{or} imaginary parts 
of the energies are degenerate, can also be defined. The analogy between EPs and 
conical intersections does not reach very far: Although the dimension of the intersection 
space is the same, most other properties are fundamentally different. In particular, a non-Hermitian 
Hamiltonian is defective at an EP, which is not the case for a Hermitian Hamiltonian at a 
conical intersection.\cite{heiss12} 

Although the role of EPs for molecular electronic resonances has not yet been investigated in 
a systematic manner,\cite{hazi83,estrada86,feuerbacher04,royal04,haxton05,benda18c} it is 
clear that they are relevant to all processes that involve two coupled resonances. This is, for 
example, the case for DEA to unsaturated halogenated compounds, which is 
presumed to lead initially to a $\pi^*$ resonance that is coupled to a $\sigma^*$ resonance 
whose PES is dissociative.\cite{fabrikant17,feuerbacher04,stricklett86,aflatooni10,benda18c} 
In analogy to bound states, non-adiabatic transitions are most likely to occur at the EP seam 
and especially near minimum-energy exceptional points (MEEPs). This motivated the 
development of an algorithm for locating MEEPs. This is again based on a 
generalization of the bound-state algorithm\cite{bearpark94} and uses the condition\cite{benda18c} 
\begin{equation} \label{eq:meep}
d/d\mathbf{R} \Big[ (E_{R1} - E_{R2} )^2 + 1/4 \cdot (\Gamma_1 - \Gamma_2)^2 \Big] = 0~,
\end{equation}
where $E_{R1}$ and $E_{R2}$ are the positions of the two resonances and $\Gamma_1$ 
and $\Gamma_2$ the corresponding widths. In addition to Eq. \eqref{eq:meep}, $E_{R1}$ 
or $E_{R2}$ needs to be minimized in the space orthogonal to the EP seam.

\begin{figure} \centering
\includegraphics[scale=0.505]{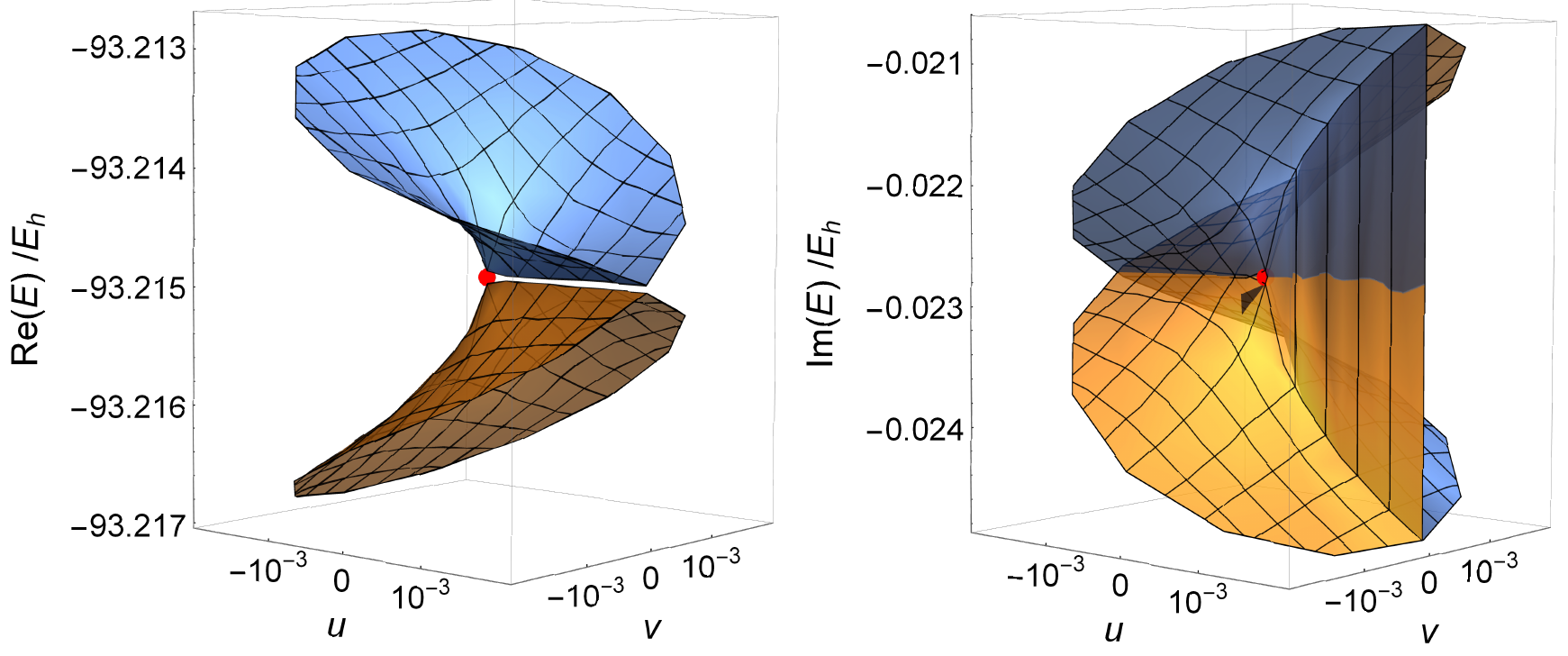}
\caption{Real (right) and imaginary (left) parts of the PES of the $^2\Pi$ and the $^2\Sigma^+$ 
resonance of HCN$^-$ computed with CAP-EOM-EA-CCSD. The EP is marked by a red dot. 
Reproduced with permission from Ref. \citenum{benda18c}.}
\label{fig:hcn}
\end{figure}

As a numerical example, Fig. \ref{fig:hcn} displays the PES of two resonances of HCN$^-$ 
near their MEEP. This shows that CAP-EOM-CCSD describes the topology of EPs consistent 
with analytical models:\cite{estrada86,feuerbacher04} Square-root energy gaps are separately 
observed for the real and imaginary parts of the energy in the branching plane. Notably, standard 
EOM-CCSD yields a flawed description of conical intersections, where the 
dimensionality of the intersection space is incorrect;\cite{koehn07,kjonstad17a,kjonstad17b} 
this difference between EOM-CCSD and CAP-EOM-CCSD can be traced back to fundamental 
differences between Hermitian and non-Hermitian operators.\cite{benda18c} 

In order to locate equilibrium structures as well as crossing seams using Eqs. \eqref{eq:mecp} 
and \eqref{eq:meep}, the first derivative of the complex resonance energy with respect to 
nuclear coordinates is required. In accordance with the interpretation of Eq. \eqref{eq:sieg1}, 
the real part and imaginary part of the gradient vector describe the change in the energy and 
decay width across the PES. They point, in general, in different directions. 

For diatomic and triatomic molecules, it is possible to evaluate the gradient vector through 
single-point energy calculations and numerical differentiation but such an approach becomes 
quickly impractical for polyatomic systems owing to increasing computational 
cost. For bound states, it was realized more than 50 years ago that an analytic evaluation of 
the energy gradient can be achieved at a cost similar to that associated with the evaluation of 
the energy itself.\cite{pulay69} Since then, analytic-derivative theory has become an important 
aspect of quantum-chemical method development and gradient expressions have been derived 
for most of the frequently used electronic-structure methods.\cite{gauss00,helgaker12} 

Corresponding developments for electronic resonances started only recently with the presentation 
of analytic gradients for CAP-HF, CAP-CCSD, and CAP-EOM-CCSD energies.\cite{benda17} 
Because all AO integrals are real-valued in CAP methods, the evaluation of the energy is easier 
here than for CBF methods. In a generic gradient expression written in the AO basis, 
\begin{equation} \label{eq:grad1}
\frac{dE}{dX} = \sum_{\mu\nu} h^X_{\mu\nu} \gamma_{\mu\nu} \; + \; 1/4 \sum_{\mu\nu\sigma\rho} 
\langle \mu\sigma || \nu\rho \rangle^X \Gamma_{\mu\sigma\nu\rho} \; + \; \sum_{\mu\nu} S^X_{\mu\nu} 
I_{\mu\nu} 
\end{equation}
with $h^X_{\mu\nu}$, $\langle \mu\nu || \sigma\rho \rangle^X$, and $S^X_{\mu\nu}$ as derivatives 
of the one-electron Hamiltonian, two-electron, and overlap integrals, it is simply necessary to 
replace the real-valued density matrices $\gamma_{\mu\nu}$, $\Gamma_{\mu\sigma\nu\rho}$, 
and $I_{\mu\nu}$ by their complex-valued counterparts for the respective CAP method. The only 
additional derivative integrals stem from the differentiation of the CAP itself and, because the CAP 
is a one-electron operator, are not relevant for the overall computational cost. It should be noted 
that these considerations only apply to the case where the CAP is included in the HF equations. 
If it is added at a later stage in a correlated calculation, additional contributions to the density 
matrices in Eq. \eqref{eq:grad1} may arise. 

A complication of CAP gradient calculations is that a molecule may be displaced relative to the 
CAP while its structure is optimized.\cite{benda17,benda18a} This can be prevented by constraining 
the gradient such that the molecule stays put relative to the origin of the CAP, which leads to a 
computationally inexpensive extra term in Eq. \eqref{eq:grad1}. The origin of the CAP can be 
chosen, for example, as center of nuclear charges. In order to deal with possible changes of the 
optimal CAP strength $\eta_\text{opt}$ across the PES, it was suggested to keep $\eta$ fixed 
during an optimization, then determine a new $\eta_\text{opt}$ according to Eq. \eqref{eq:capopt}, 
and reoptimize the molecular structure.\cite{benda17,benda18a} Typically, only 2--4 of these cycles 
are required to obtain a converged structure and CAP strength.  


\subsection{Resolution-of-the-identity approximation for electron-repulsion integrals} \label{sec:rank}
When aiming to extend the scope of a quantum-chemical method to larger systems, the 
electron-repulsion integrals (ERIs), which form a tensor of order 4, represent a major bottleneck. 
In bound-state quantum chemistry, it is common practice to exploit that the ERI tensor is 
not of full rank and several techniques have been proposed to decompose it into lower-rank 
quantities.\cite{baerends73,whitten73,dunlap79,vahtras93,beebe77,friesner85,koch03,neese09,
weigend09,hohenstein12,parrish12,izsak20} 

Only recently, however, steps were taken to extend these techniques to complex-variable methods 
for electronic resonances.\cite{hernandez19,hernandez20} Specifically, the resolution-of-the-identity 
(RI) approximation has been applied to ERIs over complex-scaled basis functions. The RI 
approximation exploits that, for a basis set of atom-centered Gaussian functions, the pair space 
of orbital products is often markedly redundant. This redundancy is particularly pronounced if large 
basis sets with many diffuse functions are used; the RI approximation thus leads to most significant 
speedups for such cases. Since complex-variable calculations often demand these 
extended basis sets, significant savings can be expected and RI methods for electronic resonances 
hold a lot of promise.

The RI approximation\cite{baerends73,whitten73,dunlap79,vahtras93} is defined according to 
\begin{equation} \label{eq:ri1}
(\mu \nu | \sigma \rho) \approx \sum_{PQ} (\mu\nu | P) [\mathbf{J}^{-1}]_{PQ} (Q | \sigma \rho) 
= \sum_{Q} B^Q_{\mu\nu} B^Q_{\sigma\rho}~,
\end{equation}
where $P$ and $Q$ refer to auxiliary Gaussian functions $\chi_P, \chi_Q$ that approximate products 
of AOs $\rho_{\mu\nu} = \chi_\mu \chi_\nu$, $J_{PQ} = \int dr_1 \int dr_2 \chi_P(r_1) r_{12}^{-1} 
\chi_Q(r_2)$, and $B_{\mu\nu}^Q = \sum_P (\mu\nu | P) [\mathbf{J}^{-1/2}]_{PQ}$. In bound-state 
quantum chemistry, Eq. \eqref{eq:ri1} is commonly applied to DFT,\cite{eichkorn95,eichkorn97} 
HF,\cite{weigend02} MP2,\cite{feyereisen93,weigend98} CC2,\cite{haettig00} and ADC(2)\cite{
haettig06} methods, where it typically entails negligible errors. Although the RI approximation does 
not reduce the formal scaling of these methods --with the notable exception of pure DFT\cite{
eichkorn95,eichkorn97}-- it does reduce absolute computation times considerably and it also 
allows to avoid storing any four-index quantity. 

\begin{figure} \centering
\includegraphics[scale=0.85]{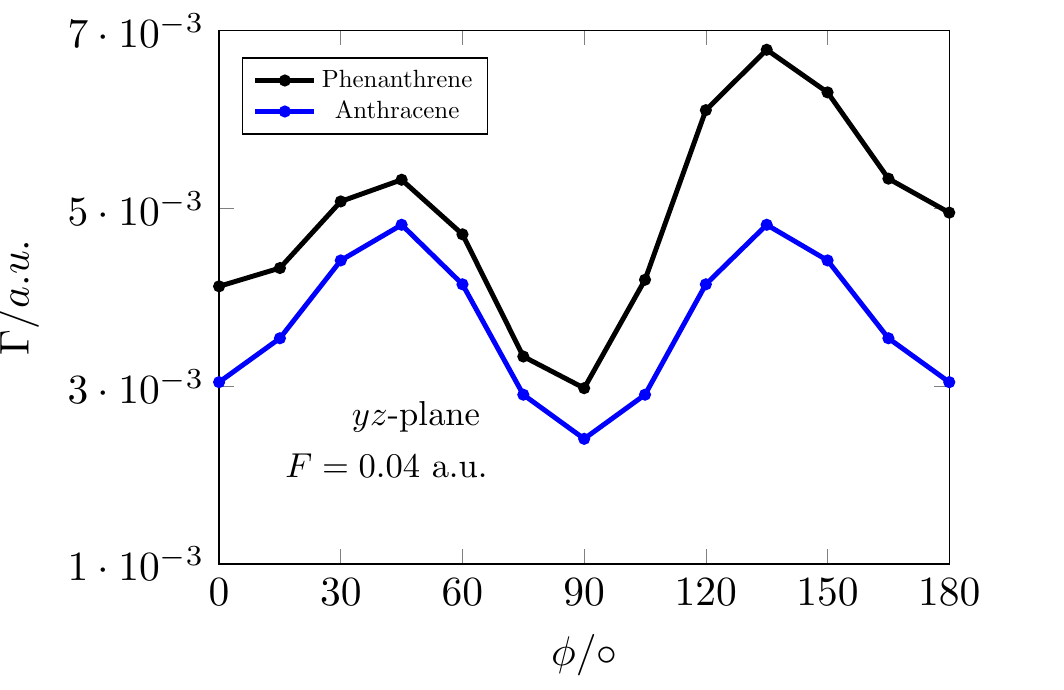}
\caption{Angle-dependent ionization rates $\Gamma$ of anthracene and phenanthrene 
(C$_{10}$H$_{14}$) in a static electric field of strength $F$=0.04 a.u. computed with RI-MP2. 
The molecules are in the $xy$ plane while the field is in the $yz$ plane and $\phi$ is the angle 
between the field and the molecular plane with $\phi=0^\circ$ corresponding to the field being 
parallel to the molecular plane. Reproduced with permission from Ref. \citenum{hernandez20}.}
\label{fig:sfi}
\end{figure}

The recent implementation of Eq. \eqref{eq:ri1} for use in RI-MP2 and RI-HF calculations 
with complex basis functions confirmed that one can indeed realize significant speedups 
in complex-variable calculations by means of the RI approximation, while 
the errors in energies and decay widths are negligible.\cite{hernandez19,hernandez20} 
CBF-RI-MP2 calculations with more than 2500 basis functions became possible, which 
enabled studying the ionization of molecules with up to ca. 50 atoms in static electric fields. 
As an example, Fig. \ref{fig:sfi} displays angle-dependent ionization rates of anthracene and 
phenanthrene.


Formally, no changes to Eq. \eqref{eq:ri1} are needed when dealing with CBFs except that 
the inversion of $\mathbf{J}$ requires care when the auxiliary basis contains CBFs as well. 
However, the derivation changes: Originally,\cite{dunlap79} Eq. \eqref{eq:ri1} was obtained 
by minimizing the functional 
\begin{equation} \label{eq:eri2}
\Delta_{\mu\nu} = \int dr_1 \int dr_2 [\rho_{\mu\nu} (r_1) - \tilde{\rho}_{\mu\nu}(r_1)] \, r_{12}^{-1} \,
[\rho_{\sigma\rho} (r_2) - \tilde{\rho}_{\sigma\rho}(r_2)]
\end{equation}
with $\tilde{\rho}_{\mu\nu}$ and $\rho_{\mu\nu}$ as approximated and exact AO products, 
respectively. This is not possible for a basis set containing CBFs because $\Delta_{\mu\nu}$ 
becomes complex as well. In Ref. \citenum{hernandez19} the absolute value $|\Delta_{\mu\nu}|$ 
was minimized instead, which also led to Eq. \eqref{eq:ri1}. 

Initial applications\cite{hernandez19} of CBF-RI-MP2 employed customized auxiliary basis sets 
including several complex-scaled shells, but it was later demonstrated\cite{hernandez20} that 
real-valued auxiliary basis sets optimized for standard RI-MP2 calculations work equally well. 
As a result, typical CBF-RI-MP2 calculations employ auxiliary basis sets that have roughly the 
same size as the original basis set. This is different from standard RI-MP2, where one usually 
uses considerably larger auxiliary basis sets. 

Very recently, Eq. \eqref{eq:ri1} has been made available for use in CC2 calculations with 
CBFs.\cite{paran22} As a multistate method, CC2 in its EOM extension can 
provide a description of a resonance together with its decay channels and is thus appropriate 
for further types of resonances besides ionization in static fields, for example, core-excited 
states and temporary anions. Since CC2 and MP2 are structurally similar,\cite{christiansen95,
haettig00} it is expected that complex-variable RI-CC2 will be applicable to resonances in 
systems with up to ca. 50 atoms as well. In addition, the RI approximation has been made 
available for CAP-MP2 and CAP-CC2 calculations. Since CAP methods are based on 
real-valued AOs, the usual RI approximation can be used and the $B$ tensors from Eq. 
\eqref{eq:ri1} become complex only upon transformation to the MO basis. 

It should be added that it is also possible to apply Eq. \eqref{eq:ri1} to the ERIs in the context 
of CCSD and EOM-CCSD.\cite{epifanovsky13} However, this reduces computation 
times and memory requirements much less than in the case of MP2 or CC2 because the 
amplitudes $t_{ij}^{ab}$ need to be stored and processed in every iteration. Given that the 
treatment of resonances often requires large basis sets, computational savings could be 
potentially higher than for bound states and complex-variable RI-CCSD may thus be a more 
viable method than its real-valued counterpart. However, no implementation has been reported 
so far.




\subsection{Quantum embedding} \label{sec:emb}
To investigate electronic resonances in complex environments, a possible further strategy 
besides the use of rank-reduction techniques consists in wave-function-theory in DFT quantum 
embedding.\cite{govind98,jones20} Here, only a small region of interest in a larger system is 
treated with a high-level method, for example EOM-CC, whereas the remainder is described 
using a lower-cost DFT approximation. Since embedding approaches rely on partitioning the 
system, they are particularly useful whenever the fragment of interest and the environment can 
be told apart easily, for example, if one deals with molecules that are surrounded by a solvation 
shell, absorbed at a surface, or enclosed in a protein coat. 

While there are many investigations where quantum embedding is used to describe properties 
and chemical reactivity of electronic ground states, applications to excitation, ionization, and 
electron attachment are a lot scarcer.\cite{izsak20} A central question is whether it is appropriate 
to use the same embedding potential for different electronic states. Recently, it was shown that 
a state-universal approach based on projection-based EOM-CCSD-in-DFT embedding\cite{
manby12,lee19,bennie17} delivers good numerical results for excited states of valence, Rydberg, 
and charge-transfer character, for valence and core ionization, and with certain reservations, also 
for electron-attached states.\cite{parravicini21} Through combination with CAPs and CBFs, the 
method has been extended to electronic resonances.\cite{parravicini21} 

As an illustration of the numerical performance, Tab. \ref{tab:emb} lists representative results for 
several types of transitions computed with embedded and regular EOM-CCSD as well as with 
DFT. In all these examples, the environment, which is treated with DFT, consists of 1--5 water 
molecules simulating microsolvation. Although it can be seen that valence excitation energies 
and valence ionization energies are well described by embedded EOM-CCSD, the usefulness 
of the approach for these transition is questionable given the good performance of DFT, represented 
in Tab. \ref{tab:emb} by the Coulomb-attenuated B3LYP density functional. 
More interesting are thus applications to core ionization and Rydberg excited states, where 
DFT struggles while embedded EOM-CCSD performs well. As concerns electron attachment, 
Tab. \ref{tab:emb} demonstrates that embedded EOM-CCSD improves on DFT but --because 
attachment energies are typically very small-- the deviation from regular EOM-CCSD is still larger 
than the actual transition energy. However, for positive attachment energies corresponding to 
temporary attachment, which can be larger, the performance of embedded CAP-EOM-CCSD 
is satisfying both for the real part of the energy and the imaginary part corresponding to the 
decay width. 

\begin{table*} \small
\caption{\ Different types of transition energies computed with regular and embedded EOM-CCSD. 
Absolute values are given for regular EOM-CCSD, deviations from these values for embedded 
EOM-CCSD. All values are in eV and have been taken from Ref. \citenum{parravicini21}}
\label{tab:emb}
\begin{tabular*}{\textwidth}{@{\extracolsep{\fill}}lllllll}
\hline
Type of state & Example & EOM-CCSD & \multicolumn{2}{c}{EOM-CCSD embedded in} & CAM-B3LYP \\
&  &  & CAM-B3LYP & PBE \\ \hline 
Valence excitation & CH$_2$O + 5 H$_2$O & \phantom{--}4.08 & \phantom{--}0.05 & \phantom{--}0.04 & --0.06 \\
Rydberg excitation & CH$_3$OH + 3 H$_2$O & \phantom{--}8.75 & --0.03 & --0.48 & --0.68 \\
Valence ionization & CH$_2$O + 5 H$_2$O & \phantom{--}11.19 & \phantom{--}0.12 & \phantom{--}0.10 & --0.15 \\
Core ionization & CH$_2$O + 5 H$_2$O & \phantom{--}540.95 & \phantom{--}0.06 & \phantom{--}0.04 & --1.40 \\ 
Electron attachment & HCF + 5 H$_2$O & --0.07 & --0.48 & --1.49 & --1.40 \\
Temporary electron attachment\footnote{Computed using CAP.} & CH$_2$O + H$_2$O & 
\phantom{--}0.88--0.10$i$ & --- & \phantom{--}0.91--0.11$i$ & --- \\
\hline
\end{tabular*}
\end{table*}

For the projection-based embedded EOM-CCSD method employed in Tab. \ref{tab:emb}, one first 
solves a standard SCF equation for the whole system, denoted A+B, using a suitable density functional. 
Based on localization of the orbitals and Mulliken population analysis, the resulting SCF wave function 
is then split into two pieces corresponding to the high-level fragment A and the environment B. 
Subsequently, a second SCF procedure is carried out with a modified Fock matrix $\tilde{\mathbf{F}}$, 
whose elements are given as
\begin{align} \label{eq:emb1}
\tilde{F}^\text{A-in-B}_{\kappa\lambda} &= \sum_{\sigma\rho} \mathcal{P}_{\kappa\mu} 
F^\text{A-in-B}_{\mu\nu} \mathcal{P}_{\nu\lambda}~, \\
F^\text{A-in-B}_{\mu\nu} &= h_{\mu\nu} + \sum_{\rho\sigma} \gamma^\text{A-in-B}_{\sigma\rho} 
[\langle \mu\sigma | \nu\rho \rangle - \langle \mu\sigma | \rho\nu \rangle] + v^\text{emb}_{\mu\nu}~.
\label{eq:emb2} \end{align}
This determines the density matrix $\boldsymbol{\gamma}^\text{A-in-B}$ that forms the basis for the 
subsequent CCSD and EOM-CCSD calculations. In Eq. \eqref{eq:emb2}, the embedding potential 
is given as $v^\text{emb}_{\mu\nu} = \sum_{\sigma\rho} [\gamma^\text{A+B}_{\sigma\rho} - 
\gamma^\text{A}_{\sigma\rho}] \, [\langle \mu\sigma | \nu\rho \rangle - \langle \mu\sigma | \rho\nu 
\rangle]$ and thus independent of $\boldsymbol{\gamma}^\text{A-in-B}$ so that it does not need 
to be recalculated during the SCF procedure. The projector $\mathcal{P}_{\mu\nu} = \delta_{\mu\nu} 
- \sum_{\rho} \gamma^\text{B}_{\mu\rho} \; S_{\rho\nu}$ removes the degrees of freedom 
corresponding to subsystem B from the variational space. This formulation of the theory 
in terms of Eqs. \eqref{eq:emb1} and \eqref{eq:emb2} is equivalent to the one originally 
introduced.\cite{manby12}

Eqs. \eqref{eq:emb1} and \eqref{eq:emb2} ensure that the SCF energy of the full system A+B is 
recovered exactly if both fragments are described at the same level of theory.\cite{manby12,lee19} 
A further advantage of projection-based embedding is that no modifications are necessary to the 
working equations of the higher-level method as long as one is only interested in the energy or 
orbital-unrelaxed properties. Likewise, quantities such as NTOs and Dyson orbitals 
that are useful to characterize excitation, ionization or electron attachment, can be evaluated in a 
straightforward manner.\cite{parravicini21} The fact that the working equations of the higher-level 
method stay the same is also the reason that the combination of projection-based embedding with 
CAP-EOM-CCSD is very easy if the CAP is active only in the EOM-CCSD calculation. In such 
an approach, the SCF calculation based on Eqs. \eqref{eq:emb1} and \eqref{eq:emb2} stays 
real-valued. The combination with CBFs is conceptually simple as well but requires the implementation 
of Eqs. \eqref{eq:emb1} and \eqref{eq:emb2} for complex numbers. 


\subsection{Partial decay widths} \label{sec:grad}
Most electronic resonances can decay into different electronic states, which are referred 
to as decay channels. For example, metastable excited states of CN$^-$ can decay into 
the $^2\Sigma^+$ and the $^2\Pi$ state of neutral CN.\cite{skomorowski18} A core-ionized 
H$_2$O molecule with electronic configuration $(1a_1)^1 (2a_1)^2 (1b_2)^2 (3a_1)^2 
(1b_1)^2$ can decay into 16 states of H$_2$O$^{2+}$ where shake-up processes and 
double Auger decay have not even been considered yet.\cite{skomorowski21b,matz21} 
In a molecule exposed to a static electric field, electrons from all orbitals can undergo 
tunnel ionization giving rise to a multitude of decay channels but the relative ease with 
which ionization from a particular orbital happens greatly depends on the orientation of 
the field and the molecule.\cite{jagau18,hernandez20}
 
Partial decay widths describing the contributions of different channels are thus of great 
importance for the chemistry and physics of electronic resonances. The notable exception 
are low-lying temporary radical anions formed by electron attachment to closed-shell 
molecules, which typically decay solely into their parent state meaning the neutral ground 
state. For all other types of resonances, partial widths are central quantities to analyze 
the fate of a metastable system. In addition, branching ratios can often be determined 
experimentally with much better accuracy than total decay widths. 

However, while there are ample theoretical data on partial decay widths of atomic resonances 
and those in diatomic or triatomic molecules, corresponding investigations of polyatomic 
molecules are comparatively scarce.\cite{manne85,zaehringer92,zaehringer92b,tarantelli94,
kolorenc11,inhester12,kolorenc20,skomorowski21,skomorowski21b,matz21} Most computations 
relied on Fano's theory\cite{fano61,feshbach62} where the determination of partial widths is 
straightforward because the total width is obtained as a sum over decay channels. This is not 
the case for complex-variable methods, where the total decay width is evaluated according to 
Eq. \eqref{eq:sieg1}, i.e., as imaginary part of an eigenvalue of the Schr\"odinger equation. 
Consequently, additional steps have to be taken to define partial decay widths and very few 
data are available so far.

In some cases, the evaluation of partial widths with complex-variable methods is facilitated by 
point-group symmetry. For example, the CAP-MRCI decay width of C$_2^{2-}$ has been 
decomposed into contributions from $\sigma_g$, $\sigma_u$, and $\pi_u$ decay channels 
by projecting the CAP onto orbitals of a particular point-group symmetry.\cite{sommerfeld00} 
In a related manner, the Auger decay width of Ne$^+$ (1s$^{-1}$) computed with CBF-CCSD 
has been decomposed into contributions from Ne$^{2+}$ states of S, P, or D symmetry by 
complex scaling only s, p, or d shells of the basis set, respectively.\cite{matz21} It has also 
been suggested to decompose the total width obtained in a CAP calculation using Fano's 
theory by considering the overlap between a Dyson orbital and a Coulomb wave representing 
the free electron\cite{gulania19} or, alternatively, by analyzing NTOs.\cite{
skomorowski18} These approaches were applied to C$_2^{2-}$ and cyanopolyyne anions 
using CCSD and EOM-EA-CCSD wave functions, respectively. 

Recently, a more general approach was introduced to evaluate partial widths in the context 
of complex-variable methods.\cite{matz21} This approach is independent of 
point-group symmetry and does not make use of Fano's theory. Instead, it is 
based on energy decomposition analysis. In initial applications, it was applied to Auger decay 
of core-ionized states described with CS-CCSD and CBF-CCSD: For a CC wave function 
describing a core-ionized state, the decay width stems solely from the imaginary part of the 
correlation energy
\begin{equation} \label{eq:pw1}
E_\text{CC} = \sum_{ijab} \Big( 1/4 t_{ij}^{ab} + 1/2 t_i^a t_j^b \Big) \langle ij || ab \rangle~,
\end{equation}
because the underlying HF reference does not capture the coupling to the continuum. It is 
thus possible to assign amplitudes $t_{ij}^{ab}$, in which $a$ or $b$ refer to the core hole, 
to a particular decay channel depending on the vacated orbitals $i$ and $j$ and to evaluate 
the partial decay width as contribution of the respective $t_{ij}^{ab}$ to $\text{Im}(E)$. If all 
amplitudes of this type are removed, a CVS-like wave function is obtained and Eq. \eqref{eq:pw1} 
yields zero for the imaginary part of the CC energy. As an alternative, the decomposition 
can be based on the CC Lagrangian
\begin{equation} \label{eq:pw2}
E_\text{CC} = \langle 0 | (1 + \Lambda) e^{-T} H e^T | 0 \rangle ~,
\end{equation}
which yields slightly different results. As a numerical example of the approach, Tab. \ref{tab:h2o} 
displays partial decay width of the 16 primary decay channels of core-ionized water computed 
with CBF-CCSD and, alternatively, based on Fano's theory. This shows overall good agreement 
between the methods with the notable exception of the $2a_12a_1$ channel, presumably because 
it has a significant admixture of other states. 

While energy expressions corresponding to other wave functions can be analyzed in a likewise 
manner, the numerical performance of the approach varies. Specifically, the 
decomposition of the EOMIP-CCSD energy in terms of $R_2$ and $L_2$ amplitudes yields 
much less reliable results because excitations that do not result in filling the core hole deliver 
unphysically large contributions to $\text{Im}(E)$. Likely, it is necessary in this case to analyze 
further components of the wave functions besides those created by $R_2$ and $L_2$. A further 
complication originates from $\text{Im}(E)$ not being exactly zero for bound states in finite basis 
sets. This is especially pronounced for CS approaches (see Sec. \ref{sec:cs2}) but does not 
impair the performance of the approach too much as Tab. \ref{tab:h2o} illustrates.

Importantly, Eqs. \eqref{eq:pw1} and \eqref{eq:pw2} apply to all complex-variable methods 
and types of resonances. The generalization from Auger decay to other processes involving 
core vacancies as well as to temporary anions and quasistatic ionization, possibly using other 
wave functions, is thus expected to be straightforward. 

\begin{table} \small
\caption{\ Partial decay widths of H$_2$O$^+$ (1s$^{-1}$) computed with different methods. 
All values in meV.}
\label{tab:h2o}
\begin{tabular*}{0.48\textwidth}{@{\extracolsep{\fill}}llll}
\hline
Decay & CBF & Fano & Fano \\
channel & CCSD\cite{matz21} & EOM-CCSD\cite{skomorowski21b} & 
MRCI\cite{inhester12,inhester14} \\ \hline
$3\text{a}_1 1\text{b}_1$ (triplet) & 0.2 & 0.5 & 0.4 \\
$1\text{b}_1 1\text{b}_1$ & 18.0 & 13.3 & 19.0 \\
$3\text{a}_1 1\text{b}_1$ (singlet) & 19.6 & 12.7 & 18.0 \\
$1\text{b}_1 1\text{b}_2$ (triplet) & 0 & 0 & 0 \\
$3\text{a}_1 3\text{a}_1$ & 12.2 & 8.9 & 13.1 \\
$1\text{b}_1 1\text{b}_2$ (singlet) & 15.7 & 10.7 & 15.2 \\
$3\text{a}_1 1\text{b}_2$ (triplet) & 0.2 & 0.4 & 0.3 \\
$3\text{a}_1 1\text{b}_2$ (singlet) & 13.4 & 9.5 & 13.2 \\
$1\text{b}_2 1\text{b}_2$ & 8.7 & 7.1 & 9.8 \\
$2\text{a}_1 1\text{b}_1$ (triplet) & 2.8 & 4.1 & 3.0 \\
$2\text{a}_1 3\text{a}_1$ (triplet) & 2.5 & 3.8 & 2.6 \\
$2\text{a}_1 1\text{b}_2$ (triplet) & 2.2 & 2.9 & 1.6 \\
$2\text{a}_1 1\text{b}_1$ (singlet) & 9.6 & 9.5 & 10.0 \\
$2\text{a}_1 3\text{a}_1$ (singlet) & 12.7 & 13.6 & 11.0 \\
$2\text{a}_1 1\text{b}_2$ (singlet) & 6.8 & 6.3 & 6.6 \\
$2\text{a}_1 2\text{a}_1$ & 21.6 & 15.3 & 4.1 \\ \hline 
All & 142.5 & 121.7 & 145.6 \\ 
\hline
\end{tabular*}
\end{table}




\section{Conclusions and outlook} \label{sec:conc}
This feature article has given an overview of the quantum chemistry of electronic resonances 
and their treatment by means of complex-variable electronic-structure methods. Because 
resonances are embedded in the continuum, their wave functions are \textit{a priori} not 
square-integrable. Complex-variable techniques afford a regularization of 
resonances and make them amenable to bound-state quantum chemistry. Quantities such 
as Dyson orbitals and natural transition orbitals can be defined and an analysis of resonances 
in terms of molecular orbital theory becomes possible.

Regularization of resonance wave functions can be achieved by means 
of complex scaling or complex absorbing potentials. While complex scaling offers several 
formal advantages and can now be applied to molecular resonances without problems using 
complex-scaled basis functions, complex absorbing potentials are more heuristic but also 
easier to integrate into existing software. In addition, CAP calculations can be sped up by 
activating the CAP only in certain steps of a calculation. However, the unphysical dependence 
of the complex resonance energy on parameters such as scaling angle and CAP strength 
presents a persistent stumbling block shared by CAP and complex-scaled approaches. 

Different types of resonances, in particular temporary anions, core-vacant states, and 
Stark resonances were discussed with respect to their electronic structure and their decay 
mechanism. These states pose distinct requirements towards the electronic-structure 
model. For temporary anions, superexcited states, core-excited and core-vacant states, 
a multistate treatment as offered by EOM-CC, SAC-CI, CC2, or ADC methods combines 
several advantages. For other states such as metastable dianions, the identification of the 
most suitable computational approach is less straightforward. The feature article has 
furthermore summarized a number of recent methodological contributions to the field: 
\begin{itemize}
\item The development of analytic gradients for CAP-CC methods has enabled the investigation 
of potential energy surfaces of polyatomic temporary anions. Special points such as equilibrium 
structures, crossing seams, and exceptional points can be located and characterized; the 
results demonstrate the relevance of such points for chemical reactions and spectroscopies 
involving electronic resonances. In addition, the availability of analytic gradients paves the 
way for conducting \textit{ab initio} molecular dynamics simulations of decaying states. Such 
calculations will be most relevant to model DEA efficiencies. 

\item The development of a resolution-of-the-identity approximation for complex basis 
functions has enabled the treatment of resonances in molecules with up to ca. 50 atoms 
at the MP2 level of theory. Because resonances require large basis sets with many diffuse 
functions, substantial speedups can be achieved by means of the RI approximation. 
Notably, standard auxiliary bases without any complex-scaled functions deliver excellent 
results for energies and decay widths. The introduction of RI methods for resonances 
can be seen as a first step towards the application of further rank-reduction techniques 
such as Cholesky decomposition.\cite{beebe77,koch03} Moreover, 
multistate methods such as complex-variable CC2 have very recently 
been combined with the RI approximation as well; these approaches are expected to 
benefit from similar advantages as MP2.

\item Projection-based wave-function-theory in density-functional-theory embedding 
provides a way to quantify the impact of the chemical environment on a resonance state. 
The method has so far only been used for microsolvated temporary anions described by 
CAP-EOM-CCSD, where is delivers good results for energies and decay widths. A corresponding 
approach based on complex basis functions that is geared towards core-vacant states has 
very recently been achieved as well. Most interesting in this context will be applications 
to decay processes such as interatomic Coulombic decay that occur only through participation 
of the environment. 

\item Partial decay widths have been made available by means of an energy decomposition 
analysis that is related to the core-valence separation. The method has been applied to 
Auger decay of core-ionized states described by complex basis functions and EOM-CCSD, 
but can be generalized to other wave functions and kinds of resonances. Partial widths 
are critically important for all states for which more than one decay channel is open 
including different types of superexcited, core-vacant, and Stark resonances. 
\end{itemize}

The present feature article illustrates that complex-variable techniques and, more general, 
quantum chemistry of electronic resonances is a field of active research. The developments 
introduced in recent years have broadened the scope of electronic-structure theory significantly 
and enabled new types of applications in computational chemistry and spectroscopy. Although 
there remain formal as well as technical issues to be solved, the implementations of complex 
scaling and complex absorbing potentials that are available today already represent a useful 
enhancement of quantum chemistry. Further research on complex-variable techniques is 
underway and provides the perspective of making these methods more practical so that, 
ultimately, they may be routinely used by non-experts as tool in quantum-chemistry program 
packages. 



\section*{Conflicts of interest}
There are no conflicts to declare.

\section*{Acknowledgements}
The author is grateful to Professors Anna I. Krylov, Nimrod Moiseyev, and Lorenz Cederbaum, 
as well as the current and former members of his research group for many fruitful discussions 
about electronic resonances and complex-variable techniques. I also thank 
Professor J\"urgen Gauss and Dr. Wojciech Skomorowski for helpful feedback on the manuscript.
Funding from the European Research Council (ERC) under the European Union's Horizon 2020 
research and innovation program (Grant Agreement No. 851766), from the German Research 
Foundation (Grant JA-2794/1-1), and from the Fonds der Chemischen Industrie is gratefully 
acknowledged. 


\renewcommand\refname{References}

\bibliography{feature_article_plain} 
\bibliographystyle{rsc} 

\end{document}